\documentclass[preprint,12pt,times]{elsarticle}

\usepackage[a4paper,left=1.2in,right=1in,top=1in,bottom=1in]{geometry}
\usepackage{amsmath,amssymb}
\usepackage{booktabs,array}
\usepackage{graphicx}
\usepackage{subcaption}
\usepackage{float}
\usepackage{siunitx}
\usepackage[svgnames]{xcolor}
\usepackage[normalem]{ulem}
\usepackage{lineno}
\usepackage[unicode=true,colorlinks=true,allcolors=blue]{hyperref}
\usepackage[nameinlink,capitalise]{cleveref}
\crefname{figure}{Fig.}{Figs.}
\Crefname{figure}{Fig.}{Figs.}

\usepackage{morefloats}


\biboptions{sort&compress}
\makeatletter\def\ps@pprintTitle{\let\@oddhead\@empty\let\@evenhead\@empty\def\@oddfoot{\hfill arXiv preprint\hfill}\let\@evenfoot\@oddfoot}\makeatother
\begin{document}

\begin{frontmatter}

\title{Interplay of interfacial energetics and selective Ni segregation in the morphology of co-precipitates in low-alloy steels}

\author[inl-cmm]{Jia-Hong Ke\corref{cor1}}
\ead{jiahong.ke@inl.gov}
\cortext[cor1]{Corresponding author.}
\author[inl-nmp]{Anshul Kamboj}
\author[inl-nmp]{Mukesh Bachhav}

\address[inl-cmm]{Computational Mechanics and Materials Department, Idaho National Laboratory, Idaho Falls, ID 83415, USA}
\address[inl-nmp]{Nuclear Materials Performance Department, Idaho National Laboratory, Idaho Falls, ID 83415, USA}

\begin{abstract}
Coupled precipitation and interfacial segregation in low-alloy steels change precipitate stability and morphology, which influence strengthening and embrittlement. Cu-rich precipitates (CRPs) and Mn--Ni--Si-rich precipitates (MNSPs) frequently form core--shell or appendage morphologies, but the interfacial energetics controlling this change are not well quantified. We combine a density-functional-theory-trained cluster expansion with lattice Monte Carlo simulations to evaluate the Fe/CRP, Fe/MNSP and Cu/MNSP interfaces within one Hamiltonian. The three coefficients based on excess internal energy follow \ensuremath{\gamma_{\rm Cu/MNSP} < \gamma_{\rm Fe/CRP} \lesssim \gamma_{\rm Fe/MNSP}} and give a positive work of adhesion and a negative spreading parameter, consistent with favorable Mn--Ni--Si attachment to a CRP and partial coverage. Only Ni segregates to the Fe/CRP boundary. Mn and Si segregate weakly on their own but become more enriched when Ni is present and co-segregate with Ni at the boundary. Ni forms an interfacial layer about one monolayer thick with an excess of \SIrange{7.4}{9.6}{atoms.nm^{-2}}, develops B2-like local order and shows an energy reduction that increases with Ni coverage. At \SI{1}{at.\%} matrix Ni, a Gibbs adsorption analysis gives a Fe/CRP free-energy reduction of about \SI{0.15}{\joule\per\square\metre}. These interface coefficients and segregation trends connect segregation, interface energy and co-precipitation and support compositional optimization in low-alloy steels.\end{abstract}


\end{frontmatter}

\section{Introduction}
\label{sec:introduction}

Co-precipitation and co-segregation can develop around existing particles in multicomponent alloys. Solutes segregate to interfaces, develop local chemical order and change interface energies, favoring attached or core--shell morphologies. In thermally aged Fe--Cu steels, atom probe tomography shows Ni and Mn segregation at coherent Fe/Cu precipitate interfaces, and the associated thermodynamic analysis reports a reduction in interfacial free energy \cite{Isheim2006}. Related segregation and co-precipitation at precipitate interfaces have been reported in aluminum alloys and Ni superalloys \cite{muddle1989precipitate,shin2017solute,poplawsky2020synergistic,biswas2011precipitates,amouyal2008segregation}. In low-alloy steels used for light water reactor pressure vessels (RPVs), this connection directly affects structural integrity. Their service life is limited by irradiation embrittlement as precipitates nucleate and coarsen \cite{odette2001embrittlement,ODETTE2019151863,odette2021precipitation,russell1984phase,xiong2024review,odette1997computational}. The main populations are coherent Cu-rich precipitates (CRPs) and Mn--Ni--Si-rich precipitates (MNSPs) \cite{odette1983dominant,miller2007embrittlement}. Because Cu has low equilibrium solubility in bcc Fe, CRPs form early at high number density, remain coherent with the bcc matrix and later transform toward the equilibrium fcc structure \cite{goodman1973fim,othen1994high,wells2014evolution,odette2009}. Ordered MNSPs develop at higher fluence and can dominate hardening, especially in steels with high Ni content \cite{sprouster2016structural,Edmondson2017,KING20181,ke2017thermodynamic,mamivand2019cumnnisi}. Their number density and size are key inputs to embrittlement models \cite{eason2013physically,ODETTE2019151863,ke2017thermodynamic,ke2022cluster,mamivand2019cumnnisi,castin2020dominant}, and these characteristics depend in part on the interface free energies that govern nucleation and coarsening.

 Atom probe studies of irradiated low-alloy steels and related ferritic steels show that CRPs and MNSPs frequently occur together. Ni, Mn and Si enrich the coherent Fe/CRP interface. The resulting particles include Cu cores partly or fully surrounded by Mn--Ni--Si shells and Cu-rich appendages attached to larger ordered Mn--Ni--Si domains \cite{styman2015characterisation,Davis2020}. In this appendage morphology, the smaller Cu-rich region becomes the attached feature once the Mn--Ni--Si domain grows larger than the CRP. This change in contact area among the CRP, MNSP and matrix affects how the co-precipitate contributes to hardening. Shell or appendage formation therefore depends on the Fe/CRP, Fe/MNSP and CRP/MNSP interface energies. Segregation states that lower these interface energies can also reduce the capillarity driving force for coarsening and help stabilize a fine precipitate dispersion.

Previous calculations have described Cu precipitation kinetics and the compositions of these particles \cite{liu1997lattice,soisson2007cu,jourdan2010cluster}. Ni segregation to the Fe/CRP interface is also well documented. Atom probe tomography (APT) measurements show positive interfacial excesses of Ni and Mn at coherent Fe/Cu precipitate interfaces and a narrow Ni-rich shell around the Cu-rich core \cite{Isheim2006}. Statistical thermodynamics and density functional theory (DFT) calculations reproduce strong Ni enrichment around curved and planar Cu-rich precipitates \cite{Gorbatov2015,Wang2019,Wang2020,Garrett2021,Garrett2022}. However, these studies do not provide the excess energies of finite particles as functions of both particle size and interfacial coverage.

Although Ni segregation to Fe/CRP interfaces is well documented, its thermodynamic role in CRP--MNSP co-precipitation remains unclear. In particular, it is not known whether Ni only accompanies MNSP formation or whether Ni adsorption changes the Fe/CRP interfacial free energy enough to alter the wetting balance.

Much less is known about the Fe/MNSP and Cu/MNSP interfaces. Shu et al.\ estimated these values by counting broken bonds between nearest neighbors in thin MnNiSi coatings on Cu cores \cite{shu2018thermodynamics}. Their pair Hamiltonian gives planar values that depend on orientation and predicts complete wetting of a CRP by MNSP. The calculation does not follow the excess energy of a finite ordered particle or include collective multicomponent segregation. It also does not separate Ni segregation from the effect of Ni coverage on Fe/CRP energy. To our knowledge, the three interfaces have not been evaluated together for finite particles with one multicomponent Hamiltonian trained with DFT data.

In this work, we focus on three related questions. The first is the balance among the Fe/CRP, Fe/MNSP and Cu/MNSP interfaces, followed by which solutes control Fe/CRP segregation energetics and how this interfacial chemistry relates to the shell-to-appendage morphology transition. We use lattice Monte Carlo simulations based on a multicomponent cluster expansion trained with DFT data for Fe alloys to examine how Ni segregation and interfacial energetics affect CRP--MNSP morphology, using the same configurational Hamiltonian as in our previous study of vacancy-stabilized MNS clustering \cite{Ke2025Vacancies}. Growth calculations provide the Fe/CRP, Fe/MNSP and Cu/MNSP interface energies as functions of particle size. Independent calculations with Ni, Mn, Si and P identify which solutes segregate to Fe/CRP and how Ni coverage changes its energy. The Ni excess is then integrated with respect to the Fe--Ni diffusion potential to calculate the relative change in Fe/CRP interfacial free energy. The resulting balance among the three interfaces is compared with the shell-to-appendage transition observed in a large simulation cell. The calculations therefore give a thermodynamic picture of how selective Ni segregation and Mn--Ni--Si co-segregation at the Fe/CRP interface connect interface energies with the CRP--MNSP morphologies observed in low-alloy steels.

\section{Methods}
\label{sec:methods}

\subsection{Cluster expansion and Monte Carlo}
\label{sec:methods_ce}

All simulations used the Fe--Ni--Mn--Si--Cr--Cu--P--vacancy cluster expansion (CE) Hamiltonian developed in our previous work \cite{Ke2025Vacancies}. The CE was trained to spin-polarized DFT energies for approximately \num{17000} relaxed configurations on the bcc lattice. The training set sampled Fe-rich and Cu-rich compositions together with B2 NiMn and the 32-site bcc mapping of G-phase Ni\(_{16}\)Mn\(_6\)Si\(_7\)Va\(_3\) \cite{whiting2020gamma2}. Details of the DFT setup and CE training procedure are reported in Ref.~\cite{Ke2025Vacancies}.

The configurational energy was expressed as \(E(\boldsymbol{\sigma})=\sum_{\alpha}V_{\alpha}\Phi_{\alpha}(\boldsymbol{\sigma})\). We used Lasso regression with L1 regularization to fit the effective cluster interactions. The cross-validation procedure and error analysis are reported in Ref.~\cite{Ke2025Vacancies}. The global tenfold cross-validation error is \SI{17.7}{\milli\electronvolt\per atom}. CASM was used for enumeration and statistical sampling \cite{puchala2023casm}. We evolved rigid bcc configurations with canonical Metropolis Monte Carlo (MC) site-exchange trials. The morphology and controlled interface analyses used two independent sets of calculations. The previously reported \(60\times60\times60\) trajectory described the change from shell to appendage at the full alloy composition \cite{Ke2025Vacancies}. Separate \(30\times30\times30\) simulations with controlled particle size and chemistry provided the three interface energies and solute effects.

\subsection{Measuring the interface energy of a growing particle}
\label{sec:methods_growth}

The excess energy was measured as one curved precipitate grew within a \(30\times30\times30\) primitive bcc lattice containing \num{27000} sites. For the Fe/CRP calculations, Cu was added one site at a time. For the MNSP calculations, each growth step added 16 Ni, 6 Mn, 7 Si and 3 vacancies, keeping the Ni\(_{16}\)Mn\(_6\)Si\(_7\)Va\(_3\) composition fixed throughout the growth path. The vacant sites were counted in \(N_{\rm C}\). Clusters were reconstructed under periodic boundary conditions using first-nearest-neighbor (1NN) connectivity. States containing a second significant cluster were excluded. The interface area was represented by an equivalent sphere with the same bcc site density,
\begin{equation}
r_{\rm eq}=\left(\frac{3N_{\rm C}}{4\pi\rho_{\rm bcc}}\right)^{1/3},
\qquad
A_{\rm eq}=4\pi r_{\rm eq}^{2},
\label{eq:area}
\end{equation}
where \(N_{\rm C}\) is the number of precipitate sites and \(\rho_{\rm bcc}\) is the site density. We use \(A\equiv A_{\rm eq}\) as the nominal interface area.

For a finite particle, the excess internal energy relative to the reference states is
\begin{equation}
E_{\rm xs}=E_{\rm cell}-N_{\rm C}g_{\rm p}
 -\sum_i\left(N_i-N_{i,\rm C}\right)\mu_i,
\label{eq:excess}
\end{equation}
Here, \(g_{\rm p}\) is the precipitate reference per site and \(N_{i,\rm C}\) is the number of sites of species \(i\) in the precipitate. The matrix reference for species \(i\) is \(\mu_i\). We used the bcc Cu endmember for the CRP and the ordered Ni\(_{16}\)Mn\(_6\)Si\(_7\)Va\(_3\) structure for the MNSP. Once these bulk contributions were removed, the excess energy was fitted as
\begin{equation}
E_{\rm xs}=bN_{\rm C}^{2/3}+cN_{\rm C}^{1/3}+d,
\qquad
\gamma_{\rm path}(N_{\rm C})=
\frac{b+c/(2N_{\rm C}^{1/3})}{k_A},
\label{eq:capillarity}
\end{equation}
where \(A_{\rm eq}=k_A N_{\rm C}^{2/3}\). The resulting \(\gamma_{\rm path}=\mathrm{d}E_{\rm xs}/\mathrm{d}A\) is the local slope along the growth path, and the constant \(d\) does not change this slope. The direct ratio, \(\gamma_{\rm direct}\equiv E_{\rm xs}/A\), was also calculated for each finite particle. Its definition and scope are summarized in Supplementary Information, Table~S1.

The same definition was used for the Fe/CRP, Fe/MNSP and Cu/MNSP interfaces. In all interface-energy calculations, each state was sampled until the formation energy converged to within \SI{3e-5}{\electronvolt} per atom. Before fitting, we averaged the simulations at each controlled size and resampled the complete growth trajectories to estimate the sampling uncertainty. The coefficients are based on excess internal energy and are averaged over orientations within the rigid lattice approximation. Fe/CRP was evaluated along growth paths at \SI{573}{\kelvin}, whereas Fe/MNSP and Cu/MNSP were evaluated at \SI{0}{\kelvin} using ordered reference states because sampling at finite temperature disorders the MNSP side at these sizes. We refer to these coefficients as interface energies. In contrast, the Gibbs adsorption analysis in \cref{sec:methods_sgc} gives a relative interfacial free-energy change at finite temperature, so the wetting quantities calculated from these coefficients include only the internal-energy contribution and cannot be used to determine an equilibrium contact angle. The measurement of \(\gamma_{\rm path}\) is illustrated in \cref{fig:method}. The numerical criteria and fit tests are given in the Supplementary Information, Section~S1. The separate Gibbs calculation is described in \cref{sec:methods_sgc} and Section~S8, and the independent flat-slab calculations in Sections~S2--S3.

\begin{figure}[ht]
\centering
\includegraphics[width=0.45\columnwidth]{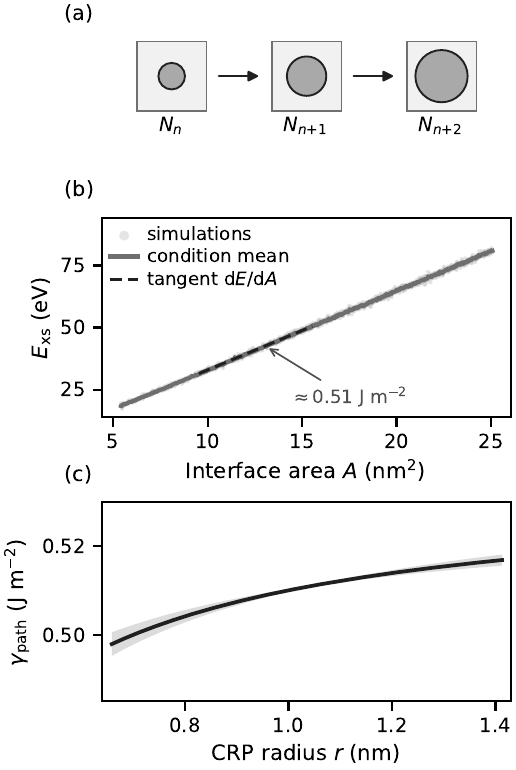}
\caption{Measurement of \(\gamma_{\rm path}\) for a curved precipitate. (a)~Schematic of the growth path, one site added per step. (b)~Excess energy from \cref{eq:excess} as a function of the area of the equivalent sphere. Points are individual simulations, the grey line is the mean at each size and the dashed line is the local tangent at \(r\approx\SI{1}{\nano\metre}\). (c)~Coefficient \(\gamma_{\rm path}(r)=\mathrm{d}E_{\rm xs}/\mathrm{d}A\) as a function of particle size. The shaded range shows the uncertainty estimated from ensemble resampling.}
\label{fig:method}
\end{figure}

The same growth procedure was repeated for Fe--Cu matrices containing one or more additional solutes at a fixed nominal composition. Mn, Ni, Si and P were tested individually, in each of the six possible pairs, and together as a Mn--Ni--Si ternary condition. In each of these eleven conditions, the named solute or solutes were held at a fixed count of 270 atoms (1~at.\% of the \num{27000} sites) for the entire growth trajectory, while Cu was added one site at a time as in the clean calculation. This fixed nominal count differs from the matrix composition measured later in the growing cell, which is lower because part of the solute segregates to the interface (\cref{sec:results_ni}). Ten independent replicas were performed for each condition at \SI{573}{\kelvin}, using the same capillarity fit, area convention and acceptance criteria as the clean baseline. These calculations provide the size-dependent coefficients and interfacial excesses in \cref{fig:systematic}, \cref{fig:specificity}, \cref{fig:cosegregation} and \cref{fig:cooperative}. The resulting coefficients are checked against independent flat-slab calculations and earlier DFT and kinetic Monte Carlo estimates for the same clean interfaces (Supplementary Information, Sections~S2--S3 and \cref{tab:shucomparison}).

\subsection{Interfacial excess and interfacial excess isotherms}
\label{sec:methods_gibbs}

For a curved CRP, the interfacial (Gibbsian) excess of an individual species depends on the choice of dividing surface \cite{krakauer1993interfacial,hellman2002measurement}. We use a Gibbs dividing surface for which the Cu excess vanishes, \(\Gamma_{\rm Cu}^{(\rm Cu)}=0\). The resulting areal excess \(\Gamma_i^{(\rm Cu)}\) is referred to as the interfacial excess throughout. For \(N\) substitutional sites, the Cu fractions in the matrix and precipitate interior are \(x_{\rm Cu}^{m}\) and \(x_{\rm Cu}^{p}\). The number of sites treated as part of the particle is \(N^{p}=(N_{\rm Cu}-x_{\rm Cu}^{m}N)/(x_{\rm Cu}^{p}-x_{\rm Cu}^{m})\). The excess of species \(i\) is then
\begin{equation}
\Gamma_i^{(\rm Cu)}=\frac{N_i-x_i^{m}(N-N^{p})-x_i^{p}N^{p}}{A_{\rm eq}}.
\label{eq:gibbs}
\end{equation}
The matrix plateau and the plateau inside the precipitate were obtained from radial composition profiles and were selected independently within each resample. This convention separates adsorption from matrix depletion and uses the same dividing surface for all solutes and particle radii.

Because particle size and Ni coverage change together during growth, additional simulations were performed to separate their effects. These calculations used eleven nominal Ni levels up to 1~at.\% and five fixed Cu counts corresponding to radii from \SI{0.69}{} to \SI{1.39}{\nano\metre}. At each size, we compared the canonical internal-energy change per equivalent area, \(\Delta\bar g_{\rm Ni}\), with an independent calculation without Ni at the same Cu count. This isolates how the energy change depends on the Ni excess at nearly fixed particle size. The complete simulation set is given in the Supplementary Information, Section~S1.

The interfacial excess isotherms were obtained from a composition sweep at a common CRP radius of \SI{1.0}{\nano\metre}. Each trajectory contributes one point at its measured matrix composition, \((x,\Gamma_i^{(\rm Cu)})\). For Ni, whose matrix content falls well below its nominal value because of strong segregation, this sweep was extended with nominal compositions as high as \num{2}~at.\%, so that the resulting matrix composition would still reach \num{1}~at.\% for the Gibbs adsorption integral endpoint (Supplementary Information, Fig.~S6c). The isotherms are described by the Langmuir--McLean form \cite{langmuir1918adsorption,mclean1957grain,hondros1977theory}
\begin{equation}
\Gamma_i(x)=\Gamma_{\rm sat}\,\frac{K\,x'}{1+K\,x'},\qquad x'=\frac{x}{1-x},
\label{eq:langmuir}
\end{equation}
with a common nominal site density \(\Gamma_{\rm sat}=\SI{13.3}{atoms.nm^{-2}}\). For the equivalent surface area of a \SI{1.0}{\nano\metre} CRP, this density corresponds to approximately 167 sites. It provides a common normalization for the coverage and was not obtained by fitting the saturation limit. From the fitted equilibrium constant \(K\) we obtain
\begin{equation}
\Delta G_{\rm seg}=-k_{\rm B}T\ln K,
\label{eq:dgseg}
\end{equation}
which is negative for favorable segregation \cite{seah1980grain,lejcek1995thermodynamics}. \(\Delta G_{\rm seg}\) is a Gibbs quantity at finite temperature, not an internal-energy change at a single particle size. Binary co-segregation was assessed at the common radius from changes in segregation free energy and interfacial excess relative to the corresponding states containing one solute \cite{guttmann1975equilibrium,guttmann1982thermodynamics,wynblatt2006anisotropy}. The effective interaction coefficient was obtained from the dependence of \(\Delta G_{\rm seg,i}-\Delta G_{\rm seg,i}^{0}\) on \(Y_j\), the interfacial coverage of the second solute \(j\) (Supplementary Information, Section~S9 and Fig.~S7).

\subsection{Fe--Ni Gibbs adsorption integral}
\label{sec:methods_sgc}

Homogeneous bcc Fe--Ni semigrand-canonical (SGC) simulations at \SI{573}{\kelvin} were used to relate the matrix composition to the Fe--Ni diffusion potential \(\Delta\mu_{\rm Ni-Fe}\). The analysis was restricted to the dilute single-phase region. At fixed temperature and fixed diffusion potentials for the other species, the Gibbs adsorption equation becomes \cite{gibbs1878equilibrium,cahn1979thermodynamics,kaplan2013review}
\begin{equation}
\Delta\gamma_{\rm Fe/CRP}(r)
=-\int_{\Delta\mu_0}^{\Delta\mu_1}
\Gamma_{\rm Ni}^{(\rm Cu)}(r,\Delta\mu)\,\mathrm{d}\Delta\mu,
\label{eq:integral}
\end{equation}
Note that this integral represents a \emph{relative} change between two matrix states. The adsorption and SGC data were jointly resampled before evaluating the integral. Further details of the resampling procedure are provided in the Supplementary Information, Sections~S1 and~S8 and Fig.~S6.

\subsection{Local order}
\label{sec:methods_order}

Local chemical order was evaluated for solute atoms attached to the CRP. The strict CRP core was defined as the largest Cu component connected through 1NN bonds. The Ni, Mn and Si atoms were also divided into components connected through 1NN bonds. A component was included in the attached population when at least one of its atoms had direct 1NN contact with the Cu core.

Bond populations were compared with random occupancy for the Ni-only, Mn--Ni, Ni--Si and Mn--Ni--Si systems. This measure uses a fixed set of interface sites that contains every bcc site outside the core and in direct 1NN contact with it. Fe sites are included, and the sites, bonds and species counts remain fixed in the random calculation. The ratios of observed to random populations compare individual bond populations with random occupancy of the same interfacial sites.

These bond counts describe only first-neighbor pairing, so on their own they do not separate B2-like order from a general preference for unlike neighbors. In a bcc lattice, B2-like order appears as a reversal between the first and second neighbor shells, with unlike Ni--Mn and Ni--Si pairs favored at 1NN and like pairs, including Ni--Ni, favored at 2NN, the same first- and second-shell Warren--Cowley reversal used to identify B2 order in other bcc alloys \cite{stana2016,kormann2021,xu2023,woodgate2025}. We combine the two shells as \(S_{\rm B2}=\tfrac{1}{2}(\rho_{\rm 2NN}-\rho_{\rm 1NN})\), where \(\rho_m\) is a Warren--Cowley-type correlation between the Ni and Mn--Si occupations in neighbor shell \(m\), calculated using the Ni fraction among the bond ends in that shell \cite{cowley1950,defontaine1994}. An equiatomic ideal B2 domain gives \(\rho_{\rm 1NN}=-1\), \(\rho_{\rm 2NN}=1\) and \(S_{\rm B2}=1\), whereas random occupation gives \(S_{\rm B2}\approx0\). At other compositions, the possible range depends on composition and on the bonds present at the interface, so we use \(S_{\rm B2}\) to measure the 1NN--2NN ordering tendency relative to random occupation. The full expression for \(\rho_m\), the 1NN and 2NN bond ratios relative to random occupancy and the random reference calculation are given in the Supplementary Information, Section~S4.

The effects of Ni, Mn and Si on local order and interfacial energy were examined at $r_{\rm eq}\approx\SI{1.0}{\nano\metre}$ with a single Cu precipitate in each cell. Total Ni was varied between 0 and 1 at.\%, and Mn and Si were either absent or present at 1 at.\%. The additional energy reduction
\begin{equation}
\begin{aligned}
 \Delta E_{\rm add}^{X}(x_{\rm Ni})={}&
 \big[E_{X+\rm Ni}(x_{\rm Ni})-E_{X+\rm Ni}(0)\big]\\
 &-\left[E_{\rm Ni}(x_{\rm Ni})-E_{\rm Ni}(0)\right],
\end{aligned}
\label{eq:additional_energy}
\end{equation}
compares the change in formation energy for $X$+Ni with that for the Ni-only system, where $X$ denotes Mn, Si or Mn--Si. Negative values mean that $X$+Ni gives a larger formation-energy reduction than the Ni-only system over the same change in total Ni concentration. Because \(\Delta E_{\rm add}^{X}\) uses formation energies for the complete simulation cell, it includes energy changes throughout the cell. Simulation and sampling details, including the random reference and the pairing of formation energies, are reported in the Supplementary Information, Section~S4 (Local order and additional energy reduction at the Fe/CRP interface).

\subsection{MNSP interfaces}
\label{sec:methods_mnsp}

Ni\(_{16}\)Mn\(_6\)Si\(_7\)Va\(_3\) on the bcc lattice was used as the ordered MNSP reference. The three vacant sites come from the 32-site bcc mapping of the G-phase structure \cite{whiting2020gamma2}. Our previous work with the same CE showed how these vacancies stabilize clusters rich in Mn, Ni and Si and change their ordering pathway \cite{Ke2025Vacancies}. The Fe/MNSP and coherent bcc Cu/MNSP calculations followed the Fe/CRP growth procedure but were evaluated at \SI{0}{\kelvin} because sampling at finite temperature readily disorders an MNSP of this size. The coherent bcc Cu/MNSP calculation represents the CRP/MNSP boundary. Its coefficient is denoted by \(\gamma_{\rm Cu/MNSP}\). The trajectory without vacancies in the large cell instead develops a phase rich in Mn, Ni and Si with order similar to B2, so we also performed an independent \SI{0}{\kelvin} calculation with stoichiometric NiMn B2 to examine how the ordered phase reference affects the wetting balance. The calculations for the two ordered phases used the same definition of excess energy and the same capillarity fit. The Fe/B2 and Cu/B2 coefficients are compared in the Supplementary Information, Section~S10 and Fig.~S8. For an MNSP \(P\) attached to a CRP \(C\) in matrix \(M\), the spreading parameter is
\begin{equation}
S_P=\gamma_{M/C}-\gamma_{C/P}-\gamma_{M/P}.
\label{eq:spreading}
\end{equation}
The total interfacial energy is
\begin{equation}
E_{\rm int}=\gamma_{M/C}A_{M/C}+\gamma_{M/P}A_{M/P}+\gamma_{C/P}A_{C/P}.
\label{eq:interface_energy_total}
\end{equation}
With these definitions, the work of adhesion is \(W_{\rm adh}=\gamma_{M/C}+\gamma_{M/P}-\gamma_{C/P}\). When all three coefficients refer to the same thermodynamic state, attachment is favored for \(W_{\rm adh}>0\), while separate precipitates are favored for \(W_{\rm adh}\leq0\). A positive spreading parameter \(S_P\geq0\) favors complete coverage of the CRP. When \(S_P<0\) and \(W_{\rm adh}>0\), attachment remains favorable but further spreading raises the energy. A finite cap or appendage is then preferred \cite{smith1948grains}.

\subsection{Morphology}
\label{sec:methods_morphology}

The morphology was quantified from the previously reported \(60\times60\times60\) canonical MC trajectory for Fe--0.2Cu--1.5Ni--1.5Mn--0.5Si--0.1P at.\% \cite{Ke2025Vacancies}. Precipitates containing at least \num{80} atoms were identified using a first-nearest-neighbor (1NN) cutoff of \SI{0.25}{\nano\metre}. The displacement of the Cu center was normalized by the radius of the attached MNSP, giving the ratio \(d_{\rm Cu}/r_{\rm MNSP}\) used throughout this work. A value of zero represents concentric coverage. A value close to unity places the Cu center near the edge of the MNSP domain. The sampled configurations cover \num{20}--\num{5000} Monte Carlo passes and were rendered with OVITO \cite{Stukowski2010OVITO}. One Monte Carlo pass contains \(N\) attempted exchanges between lattice sites, where \(N\) is the number of active sites in the simulation cell. Clusters and interfaces were tracked along this trajectory to measure center displacement, boundary contact fractions and interface areas.

\section{Results}
\label{sec:results}

\subsection{Evolution from shell to appendage}
\label{sec:results_morphology}

\begin{figure*}[ht]
\centering
\includegraphics[width=0.85\textwidth]{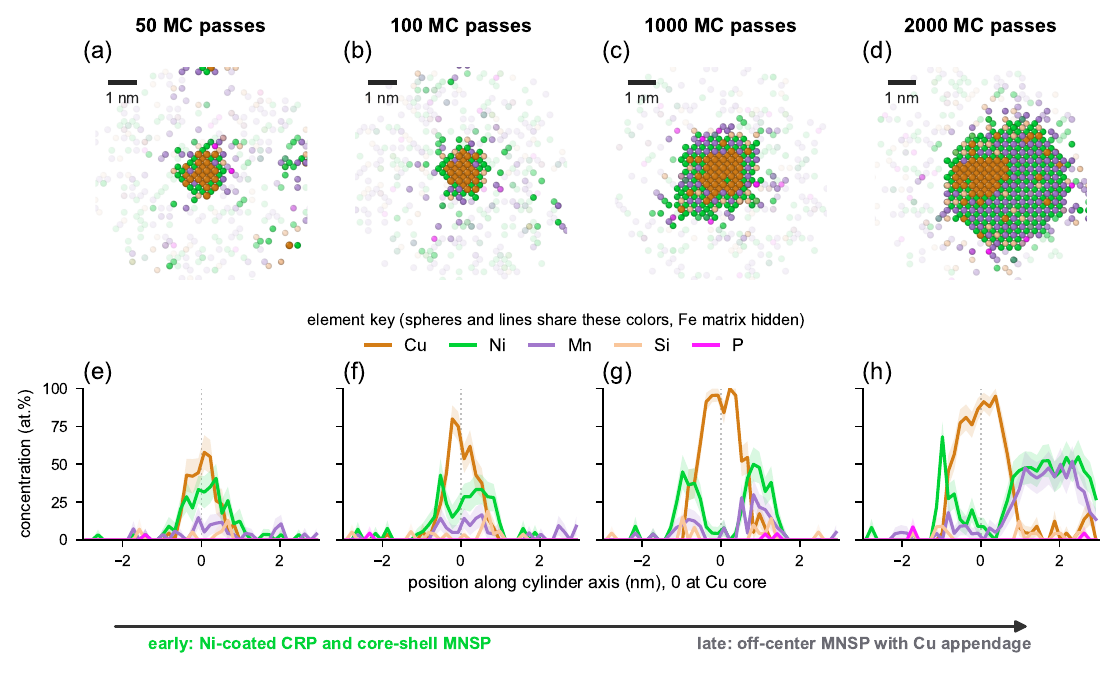}
\caption{Evolution from a shell enriched in Ni to an asymmetric CRP--MNSP appendage in Fe--0.2Cu--1.5Ni--1.5Mn--0.5Si--0.1P at.\% without vacancies at \SI{573}{\kelvin}. The Monte Carlo trajectory was reported in Ref.~\cite{Ke2025Vacancies} and is reanalyzed here. Central cross sections and one-dimensional concentration profiles along a cylinder axis through the Cu core are shown at \num{50}, \num{100}, \num{1000} and \num{2000} MC passes, and the final state retains a small exposed Fe/CRP boundary. The labels early and late in the figure refer to the order of MC passes, not to physical time.}
\label{fig:morphology}
\end{figure*}

A large-cell Monte Carlo trajectory follows how a Cu-rich precipitate and its attached Mn--Ni--Si phase develop from an initially homogeneous solid solution. In this trajectory, the canonical Monte Carlo calculation begins from the full alloy composition \cite{Ke2025Vacancies}. \Cref{fig:morphology} shows the corresponding evolution from a shell to an appendage. A shell enriched in Ni first surrounds a small Cu-rich region (\cref{fig:morphology}a). As Ni accumulates, the shell thickens (\cref{fig:morphology}b), closes into an ordered core--shell particle (\cref{fig:morphology}c) and finally becomes an asymmetric appendage attached to a larger domain rich in Mn, Ni and Si (\cref{fig:morphology}d). This sequence is consistent with atom probe observations in irradiated low-alloy steels and related ferritic steels \cite{Isheim2006,styman2015characterisation,wells2014evolution,Davis2020,kamboj2024dose}.

Cylinder 1D concentration profiles in \cref{fig:morphology}e--h trace the same change along the axis from the Cu core to the MNSP. At \num{50} MC passes, a narrow Cu peak is centered within a broader region enriched in Ni, while the Mn and Si enrichments remain weak. By \num{100} MC passes, Ni, Mn and Si have increased on both sides of the sharper Cu peak to form a core--shell profile. The Mn--Ni--Si enrichment shifts mainly to one side by \num{1000} MC passes. At \num{2000} MC passes, the Cu-rich peak is attached to a broad domain rich in Mn, Ni and Si, and Ni remains enriched at the exposed Fe/CRP boundary on the opposite side.

\begin{figure*}[ht]
\centering
\includegraphics[width=0.75\textwidth]{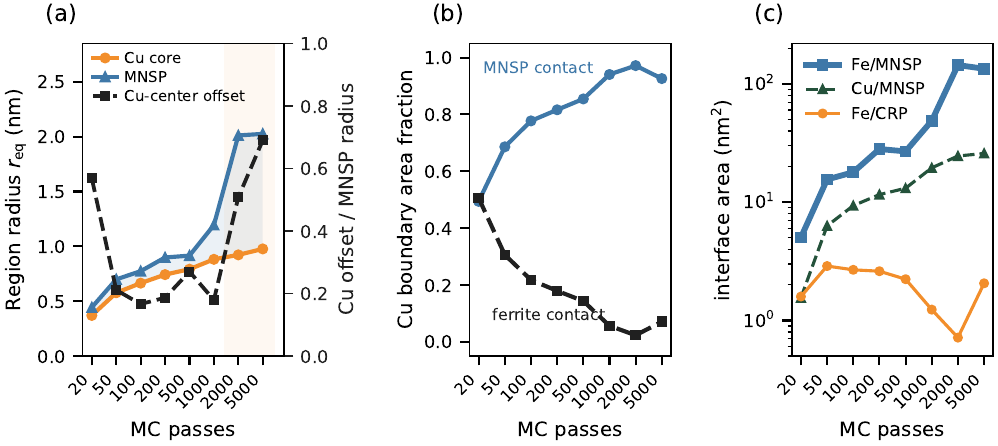}
\caption{Morphology measures for the trajectory in \cref{fig:morphology}. (a)~Equivalent radii of the Cu core and the MNSP (left axis) and normalized displacement of the Cu center, \(d_{\rm Cu}/r_{\rm MNSP}\) (right axis), as functions of MC passes. (b)~Fractions of the Cu boundary in contact with the MNSP and ferrite. (c)~Areas of the Fe/CRP, Fe/MNSP and Cu/MNSP interfaces.}
\label{fig:morphologyquant}
\end{figure*}

 Figure~\ref{fig:morphologyquant} summarizes the shell-to-appendage transition. As the MNSP grows from a thin shell into a larger ordered phase, the Cu center moves toward its edge (\cref{fig:morphologyquant}a). Over the same interval, the MNSP contact fraction increases and the ferrite contact falls to a small fraction (\cref{fig:morphologyquant}b). Corresponding areas of the Fe/CRP, Fe/MNSP and Cu/MNSP interfaces for the same trajectory are shown in \cref{fig:morphologyquant}c. In the final state, only a small Fe/CRP area remains exposed and the geometry appears as a Cu-rich appendage once the MNSP becomes larger than the CRP.

An independent calculation with structural vacancies also develops an asymmetric appendage. The Mn, Ni, Si and vacancy populations follow the Ni\(_{16}\)Mn\(_6\)Si\(_7\)Va\(_3\) site ratio used in our previous work \cite{Ke2025Vacancies}. At the same MC pass, this calculation has a Cu-rich core and a larger displacement from the center than the calculation without vacancies. It does not form the closed core--shell stage found without vacancies. Despite their different early configurations, both calculations develop asymmetric co-precipitates.

\subsection{Clean Fe/CRP, Fe/MNSP, and Cu/MNSP interfaces}
\label{sec:results_interfaces}

To provide a reference, we first evaluate the coherent Fe/CRP boundary. Its excess energy increases nearly linearly with equivalent interface area over the range containing a single cluster. The ten simulations show a narrow spread at most particle sizes (\cref{fig:method}b,c). From \(r_{\rm eq}\approx\SI{0.66}{}\) to \SI{1.41}{\nano\metre}, the coefficient remains between \SI{0.50}{} and \SI{0.52}{\joule\per\square\metre}. This range overlaps the relaxed planar DFT range of \SIrange{0.45}{0.61}{\joule\per\square\metre} \cite{Wang2020}. It is higher than the \SIrange{0.30}{0.34}{\joule\per\square\metre} reported by Seko et al.\ for small clusters \cite{Seko2004}. Further numerical sensitivity tests and alternative particle-size treatments are summarized in the Supplementary Information, Section~S1.

\begin{figure*}[ht]
\centering
\includegraphics[width=0.65\textwidth]{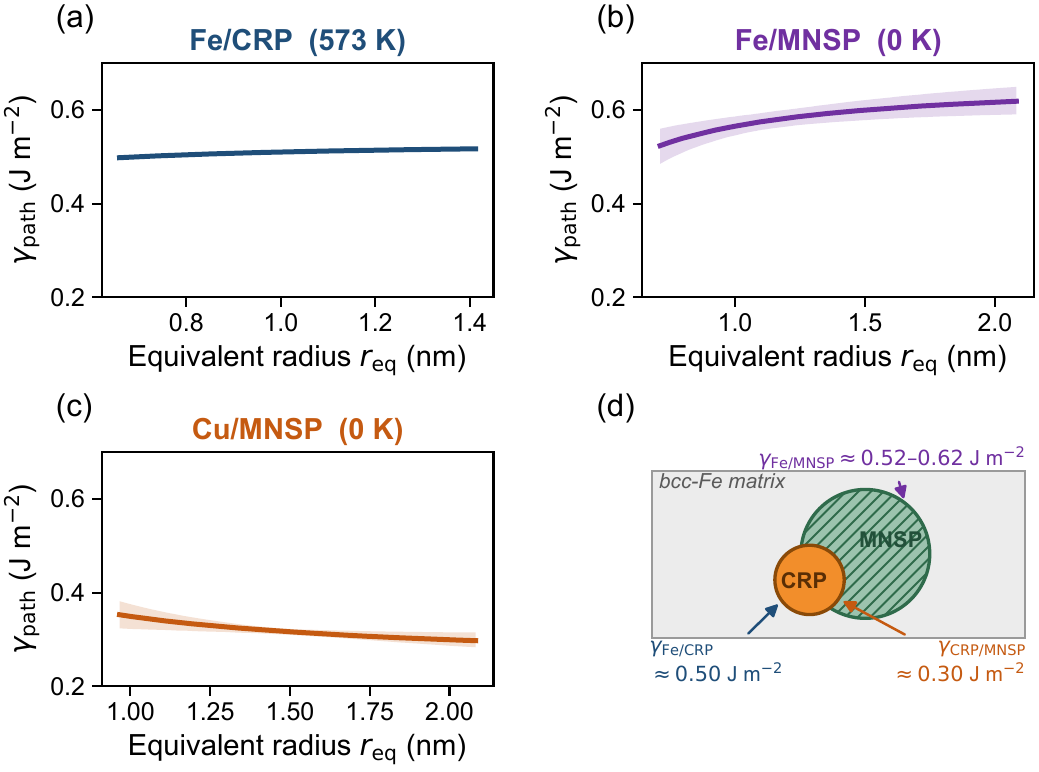}
\caption{Comparison of the coefficients based on internal energy for CRP--MNSP co-precipitation. (a--c)~Coefficients \(\gamma_{\rm path}\) for Fe/CRP, Fe/MNSP and Cu/MNSP with the same vertical range. (d)~Schematic of attachment with partial wetting for \(W_{\mathrm{adh},E}>0\) and \(S_E<0\), where \(\gamma_{\rm CRP/MNSP}\) denotes the Cu/MNSP coefficient. The geometry is illustrative because the coefficients were evaluated at different temperatures.}
\label{fig:threeinterface}
\end{figure*}

The \SI{0}{\kelvin} growth calculations give an Fe/MNSP coefficient between \SI{0.52}{} and \SI{0.62}{\joule\per\square\metre}, which is at or slightly above the clean Fe/CRP value. We used \SI{0}{\kelvin} because finite-temperature sampling readily disorders an MNSP of this size. The Cu/MNSP coefficient is lower and ranges from \SI{0.30}{} to \SI{0.36}{\joule\per\square\metre} at the same temperature. The three interfaces therefore follow \(\gamma_{\rm Cu/MNSP}<\gamma_{\rm Fe/CRP}\lesssim\gamma_{\rm Fe/MNSP}\) (\cref{fig:threeinterface}). Over their common size interval, this ranking leads to \(W_{\mathrm{adh},E}>0\) and \(S_E<0\), consistent with attachment and partial wetting when the coefficients based on internal energy are compared. The size dependence of the three coefficients is shown in \cref{fig:threeinterface}a--c, and the corresponding partial-wetting geometry is shown in \cref{fig:threeinterface}d. Flat-slab Fe/CRP and Fe/MNSP interfaces (\(\gamma_{\rm flat}\), Supplementary Information, Tables~S2--S3) agree with \(\gamma_{\rm path}\) in both trend and magnitude. For Cu/MNSP, the flat-slab coefficients are higher and overlap the upper end of the \(\gamma_{\rm path}\) range. The direct ratio \(\gamma_{\rm direct}\equiv E_{\rm xs}/A\) offers a second check on \(\gamma_{\rm path}\) for the same three interfaces. It agrees closely with \(\gamma_{\rm path}\) for Fe/CRP (\SIrange{0.519}{0.522}{\joule\per\square\metre}) and Fe/MNSP (\SI{0.579}{\joule\per\square\metre}), and is close to the low end of the Cu/MNSP range (\SI{0.318}{\joule\per\square\metre}). For Fe/MNSP and Cu/MNSP, \(\gamma_{\rm direct}\) is averaged over the largest size simulations.

\subsection{Individual-solute segregation and Fe/CRP free energy}
\label{sec:results_ni}

\begin{figure*}[ht]
\centering
\includegraphics[width=0.90\textwidth]{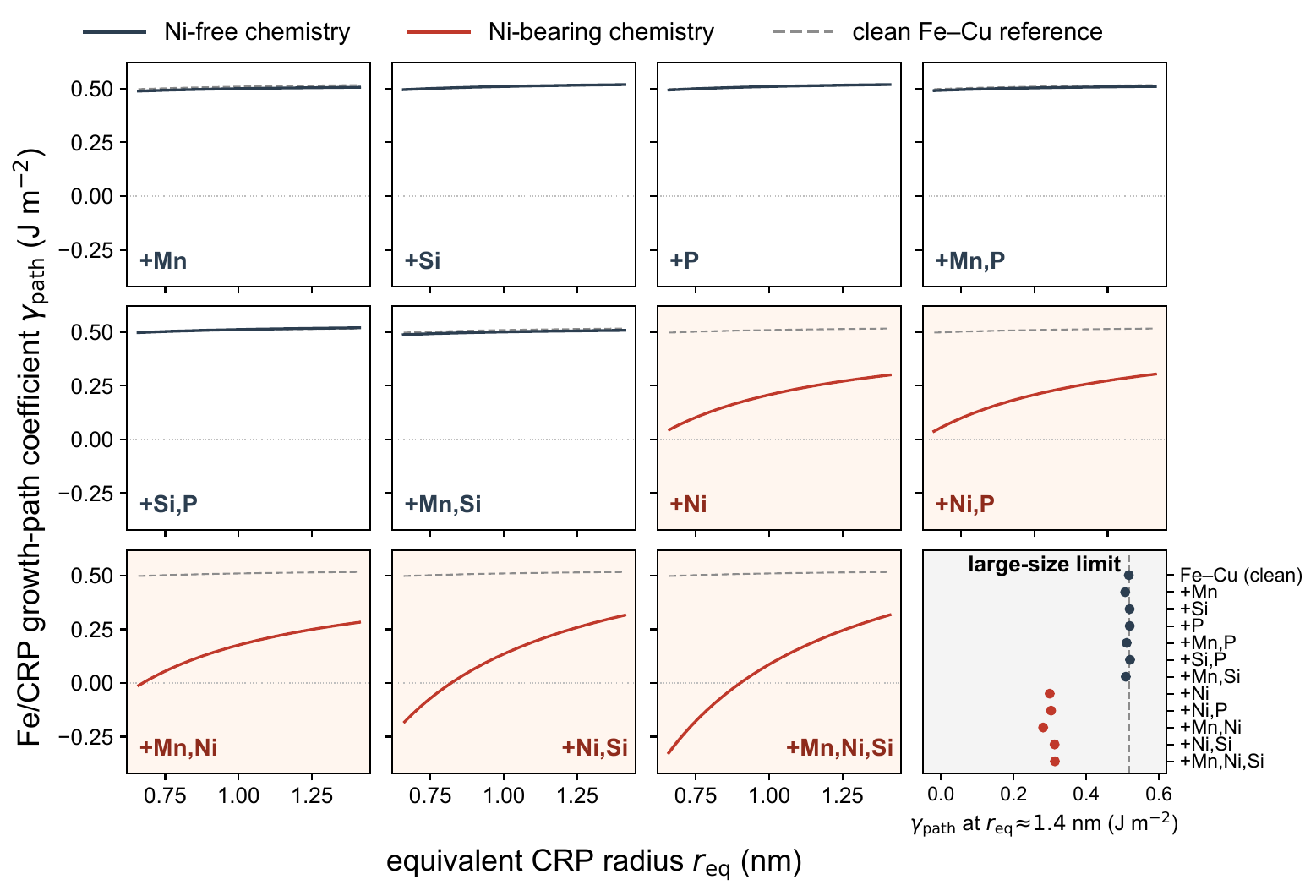}
\caption{Dependence of the Fe/CRP coefficient \(\gamma_{\rm path}(r_{\rm eq})\) on size and solute chemistry at \SI{573}{\kelvin}. Each subfigure gives an independent condition containing 1~at.\% of the named solute or solutes. The dashed grey curve is the clean Fe--Cu reference and the chemistries containing Ni are shaded. Conditions without Ni remain near the reference, while those with Ni are lower at small size and approach the reference as the particle grows. At the smallest simulated sizes, \(\gamma_{\rm path}\) becomes negative for the Ni,Si and Mn,Ni,Si conditions and is near zero for Mn,Ni.}
\label{fig:systematic}
\end{figure*}

To identify which solutes modify the Fe/CRP boundary, independent growth calculations were performed for Fe--Cu matrices containing 1~at.\% Ni, Mn, Si or P. Each condition was compared with the clean Fe--Cu reference. Mn, Si and P have little effect on the coefficient over the sampled size range, whereas Ni lowers it throughout the growth calculation.  The complete growth calculations with one or several solutes are compared in \cref{fig:systematic}, and the role of Ni coverage is examined below.

\begin{figure*}[ht]
\centering
\includegraphics[width=0.7\textwidth]{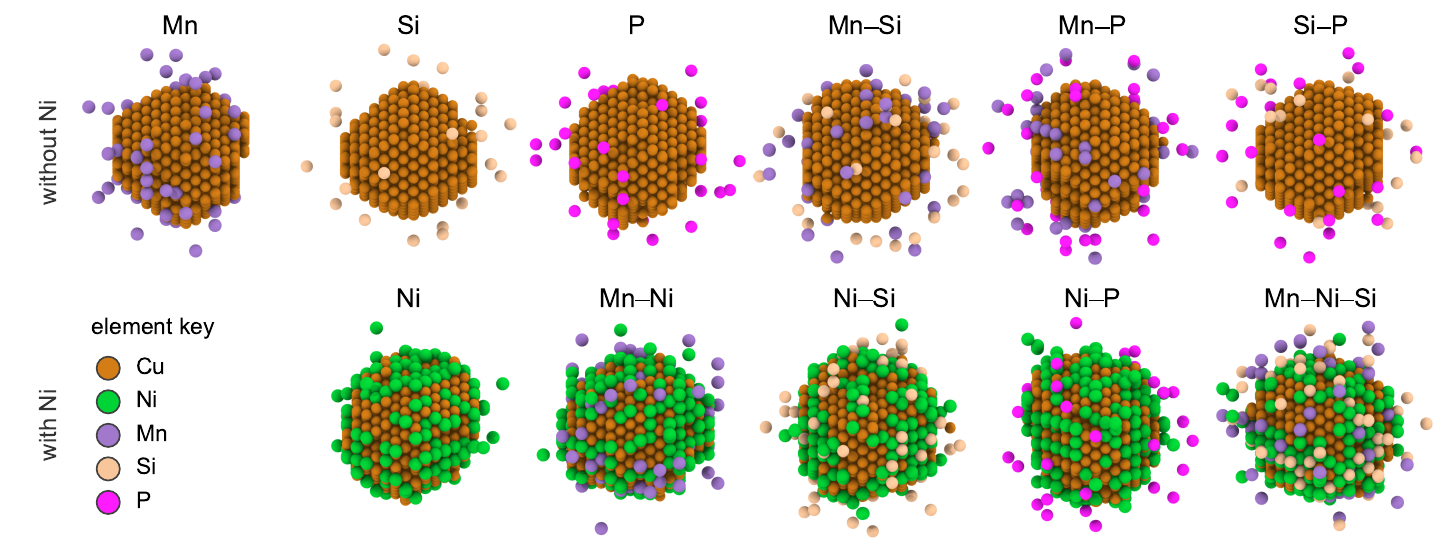}
\caption{Solute enrichment at the Fe/CRP interface at \SI{573}{\kelvin} for a fixed Cu size. Conditions without Ni are in the top row and conditions with Ni in the bottom row. The Fe matrix is hidden and the colors follow the key.}
\label{fig:crpsnapshots}
\end{figure*}

\begin{figure*}[ht]
\centering
\includegraphics[width=0.75\textwidth]{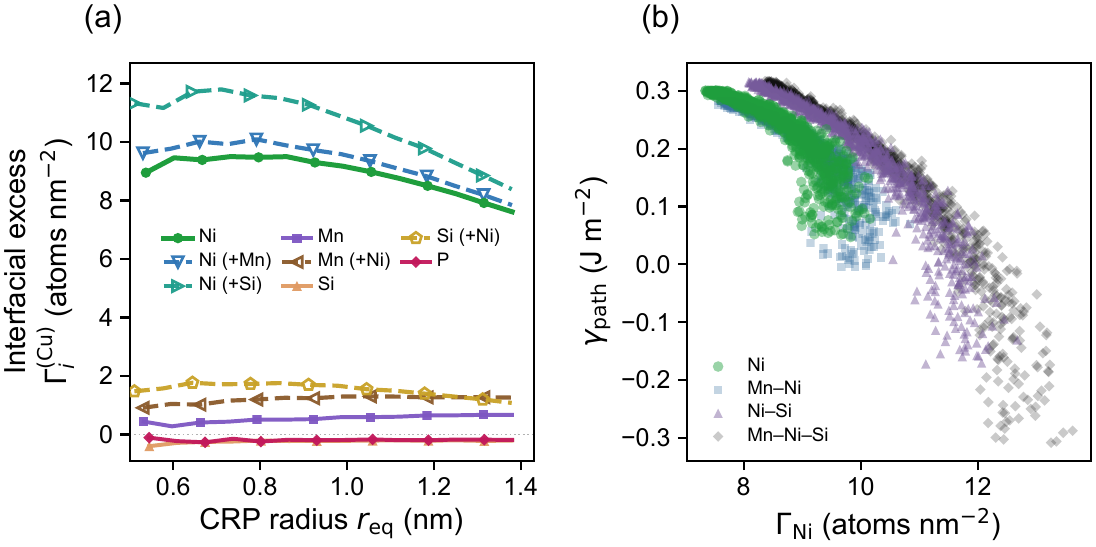}
\caption{Segregation and energetics at the Fe/CRP interface at \SI{573}{\kelvin}. (a)~Interfacial excess \(\Gamma_i^{(\rm Cu)}(r_{\rm eq})\) as a function of particle size for each solute and multisolute condition.  (b)~\(\gamma_{\rm path}\), evaluated at each condition's particle size, as a function of Ni excess.}
\label{fig:specificity}
\end{figure*}

At 1~at.\% solute, Ni lowers the coefficient by more than 50\% relative to the clean Fe/CRP value at the smallest simulated sizes, while the Mn, Si and P coefficients remain close to the clean reference. Configurations also show a dense interfacial shell only when Ni is present (\cref{fig:crpsnapshots}). For the Ni,Si and Mn,Ni,Si conditions, \(\gamma_{\rm path}\) becomes negative at the smallest simulated sizes (\cref{fig:systematic}). Here, a negative \(\gamma_{\rm path}\) denotes a negative local slope \(\mathrm{d}E_{\rm xs}/\mathrm{d}A\) along the fixed-composition growth path, not a negative absolute interfacial free energy. At these sizes the Ni excess per unit interface area is near its maximum, and \(\gamma_{\rm path}\) approaches the clean reference as this excess decreases with growth. The corresponding interfacial excesses in \cref{fig:specificity} show how solute accumulation relates to the energy change.

Ni has an interfacial excess in \cref{fig:specificity}a that is approximately one order of magnitude greater than the excesses of Mn, Si and P. Mn segregates only weakly when simulated by itself, while Si and P are slightly depleted over most of the size range. In the presence of Ni, both Mn and Si become enriched.

To separate Ni coverage from particle size, we used growth calculations with fixed Cu counts. For all chemistries containing Ni, \(\gamma_{\rm path}\) decreases as the Ni excess increases (\cref{fig:specificity}b). This relation applies to Ni by itself and to Ni with Mn, Si or both. The energy reduction therefore follows the Ni coverage. At the highest nominal Ni content, the interfacial excess is between \SI{7.4}{} and \SI{9.6}{atoms.nm^{-2}} for the five particle sizes (matrix Ni content \SIrange{0.35}{0.79}{at.\%}, depending on particle size) and is comparable to a dense substitutional monolayer. During growth, the excess first increases and then decreases because the available Ni is distributed over a larger interface. This change in coverage produces the apparent size dependence of the coefficient in \cref{fig:systematic}.

\begin{figure*}[ht]
\centering
\includegraphics[width=0.70\textwidth]{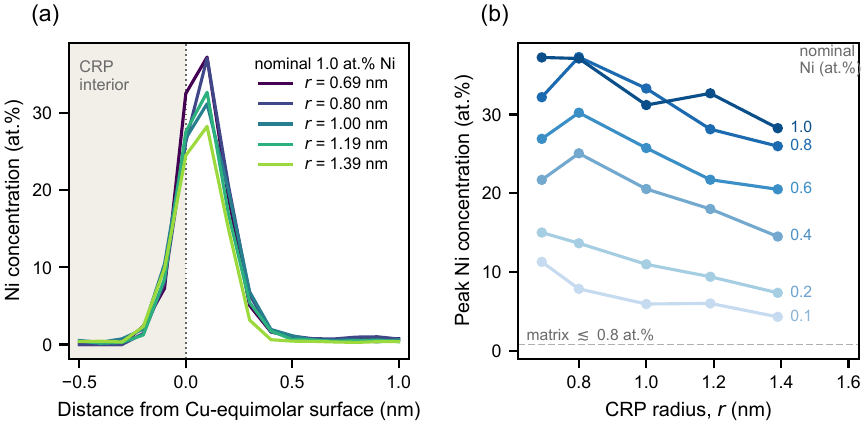}
\caption{Radial Ni concentration profile across the Fe/CRP interface at \SI{573}{\kelvin}. (a)~Ni concentration as a function of distance from the CRP boundary. The sharp interfacial peak has a width comparable to a monolayer. (b)~Dependence of the peak height on nominal Ni content and particle size.}
\label{fig:radialprofile}
\end{figure*}

Radial profiles show that Ni forms a sharp interfacial peak rather than a diffuse concentration gradient (\cref{fig:radialprofile}). The peak remains approximately one monolayer wide as the nominal Ni content changes. Its enrichment is confined to about \SI{0.3}{\nano\metre} around the CRP boundary, and its height increases as the nominal Ni content increases or the particle size decreases. Even at the lowest nominal Ni content, the interfacial concentration exceeds the matrix concentration by more than two orders of magnitude (\cref{fig:radialprofile}a). Figure~\ref{fig:radialprofile}b shows how the peak height depends on nominal Ni content and particle size. This narrow Ni-enriched shell around the Cu-rich region is consistent with atom probe measurements that show Ni enrichment at coherent Fe/Cu interfaces in low-alloy steels \cite{Isheim2006,styman2015characterisation}.

\begin{figure*}[ht]
\centering
\includegraphics[width=0.9\textwidth]{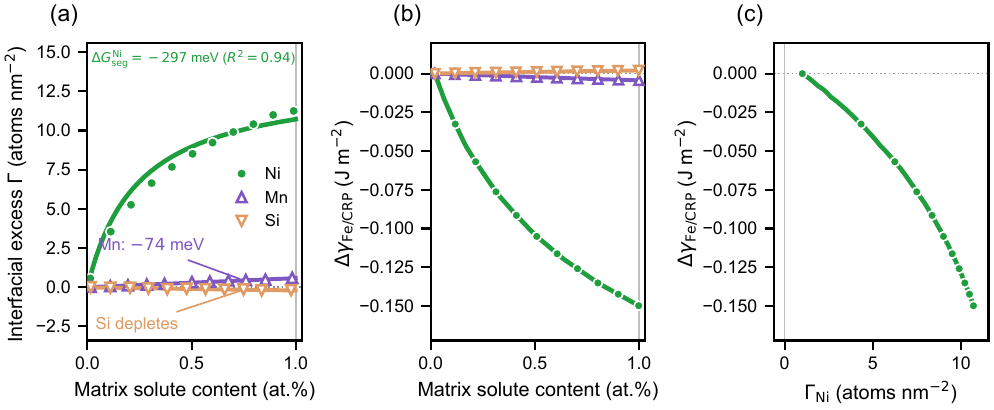}
\caption{Segregation of individual solutes and Fe/CRP free energy at \SI{573}{\kelvin} and a common \SI{1.0}{\nano\metre} radius. (a)~Langmuir--McLean isotherms for Ni, Mn and Si up to \num{1}~at.\% matrix solute content. (b)~Change in free energy from the Gibbs adsorption integral. (c)~Gibbs adsorption result for Ni as a function of interfacial excess.}
\label{fig:isotherm}
\end{figure*}

Segregation at a particle radius of \SI{1.0}{\nano\metre} was measured as a function of the matrix composition far from the interface.  At this fixed radius, the matrix Ni, Mn or Si content was raised from zero to obtain the matrix composition \(x_i^{m}\) and the corresponding interfacial excess \(\Gamma_i^{(\rm Cu)}\) (\cref{fig:isotherm}a). \(x_i^{m}\) is lower than the nominal content imposed on the simulation cell because part of the added solute segregates to the interface. The Langmuir--McLean form describes the Ni data well and gives \(\Delta G_{\rm seg}=\SI{-297}{\milli\electronvolt}\). The Mn enrichment is weaker and deviates from the ideal form with one type of site. We therefore do not report one value of \(\Delta G_{\rm seg}\) for Mn that is independent of composition. Si remains depleted across the sampled range and no McLean fit is reported.

The Ni isotherm in \cref{fig:isotherm}a and the semigrand-canonical Fe--Ni relation in Supplementary Information, Fig.~S6a, together provide the two quantities used in the Gibbs adsorption integral. The measured matrix interval, from near \num{0} to \num{1}~at.\% Ni, was mapped onto the diffusion potential \(\Delta\mu_{\rm Ni-Fe}\). Integrating \(-\Gamma_{\rm Ni}^{(\rm Cu)}\,\mathrm{d}\Delta\mu_{\rm Ni-Fe}\) gives \(\Delta\gamma_{\rm Fe/CRP}\) relative to the endpoint with lower Ni content. Mn and Si were each integrated the same way, along their own diffusion-potential path.

 At a fixed radius of \SI{1.0}{\nano\metre} and the largest simulated matrix content of \num{1}~at.\%, the same Gibbs integral gives \(\Delta\gamma_{\rm Fe/CRP}=\SI{-0.150}{\joule\per\square\metre}\) for Ni (\(\Gamma_{\rm Ni}=\SI{10.72}{atoms.nm^{-2}}\)), \SI{-0.004}{\joule\per\square\metre} for Mn (\(\Gamma_{\rm Mn}=\SI{0.58}{atoms.nm^{-2}}\)) and \SI{+0.002}{\joule\per\square\metre} for Si (\(\Gamma_{\rm Si}=\SI{-0.21}{atoms.nm^{-2}}\)) (Supplementary Information, Fig.~S6c). Ni therefore gives the dominant Gibbs contribution over this wider composition range. These Gibbs changes are smaller than the canonical internal-energy reductions reported above because the two calculations follow different thermodynamic paths.

\subsection{Co-segregation at the Fe/CRP interface}
\label{sec:results_coseg}

\begin{figure*}[ht]
\centering
\includegraphics[width=0.75\textwidth]{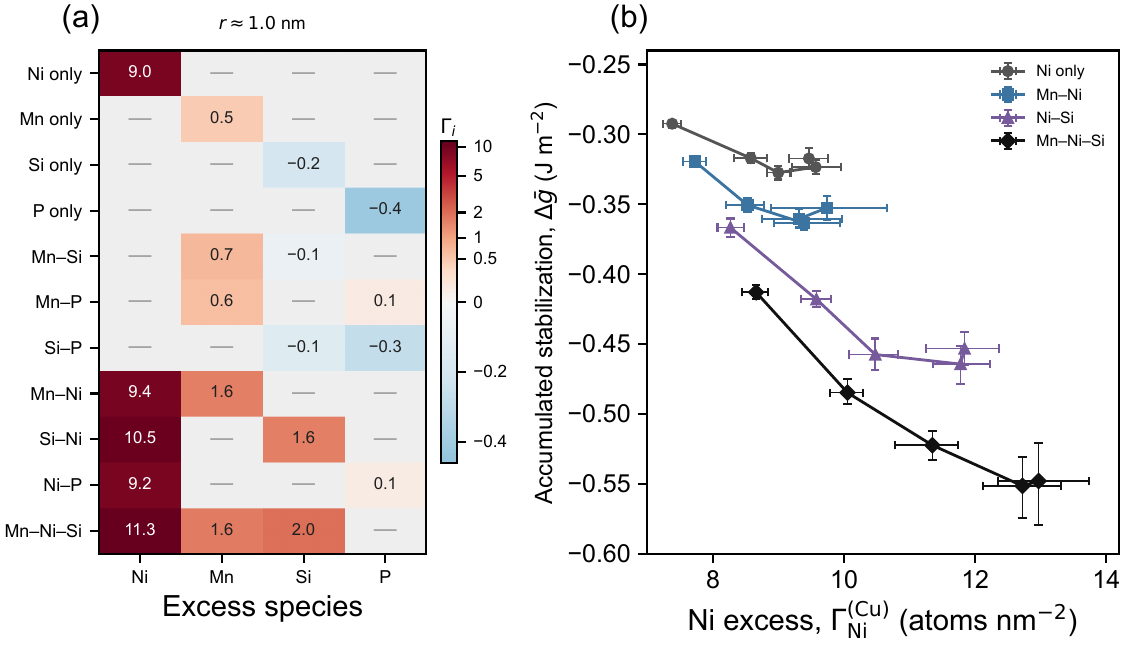}
\caption{Co-segregation at the Fe/CRP interface at \SI{573}{\kelvin} and \(r\approx\SI{1.0}{\nano\metre}\). (a)~Interfacial excesses for compositions with one, two or three solutes. (b)~Accumulated internal-energy reduction as a function of Ni excess, with the order Mn--Ni--Si \(>\) Ni--Si \(>\) Mn--Ni \(>\) Ni.}
\label{fig:cosegregation}
\end{figure*}

To examine co-segregation, we performed binary growth calculations to see how Ni changes the Mn, Si and P excesses at the Fe/CRP boundary. At \(r_{\rm eq}\approx\SI{1.0}{\nano\metre}\), Ni increases the Mn excess while the Ni excess itself changes only weakly over this range (\cref{fig:cosegregation}). Ni and Si increase one another's excess and produce the strongest reciprocal co-segregation. These trends remain from approximately \SI{0.7}{} to  \SI{1.4}{\nano\metre}. Both pairs containing Ni also reduce \(\gamma_{\rm path}\) below the value for the Ni-only system. The interfacial excesses for the one-, two- and three-solute conditions are shown in \cref{fig:cosegregation}a, and the accumulated internal-energy reduction is shown in \cref{fig:cosegregation}b.

P develops a finite boundary excess but contributes little additional energy reduction. Ni--P and Mn--P increase the P excess without a clear reciprocal change in the other solute, while the Si--P condition changes neither excess. The Mn--Si pair without Ni also remains near the clean reference and shows no detectable co-segregation. Every binary condition with a strongly reduced coefficient therefore contains Ni. The configurations in \cref{fig:crpsnapshots} also show that only a small amount of P enters the dense shell enriched in Ni \cite{kamboj2023effect}.

The Mn--Ni--Si calculation shows how Mn and Si together change the Ni segregation free energy. Mn and Si each make Ni segregation more favorable.  The reduction in internal energy follows Mn--Ni--Si \(>\) Ni--Si \(>\) Mn--Ni \(>\) Ni at every common radius (\cref{fig:cosegregation}b). At \(r\approx\SI{1.0}{\nano\metre}\), \(\Delta\bar g\) reaches \SI{-0.522}{\joule\per\square\metre} for the ternary condition, compared with \SI{-0.458}{}, \SI{-0.364}{} and \SI{-0.327}{\joule\per\square\metre} for Ni--Si, Mn--Ni and the Ni-only system. Because these are canonical energy changes, they are not combined with the Gibbs result into a single multicomponent \(\Delta\gamma\).

\begin{figure*}[ht]
\centering
\includegraphics[width=0.82\textwidth]{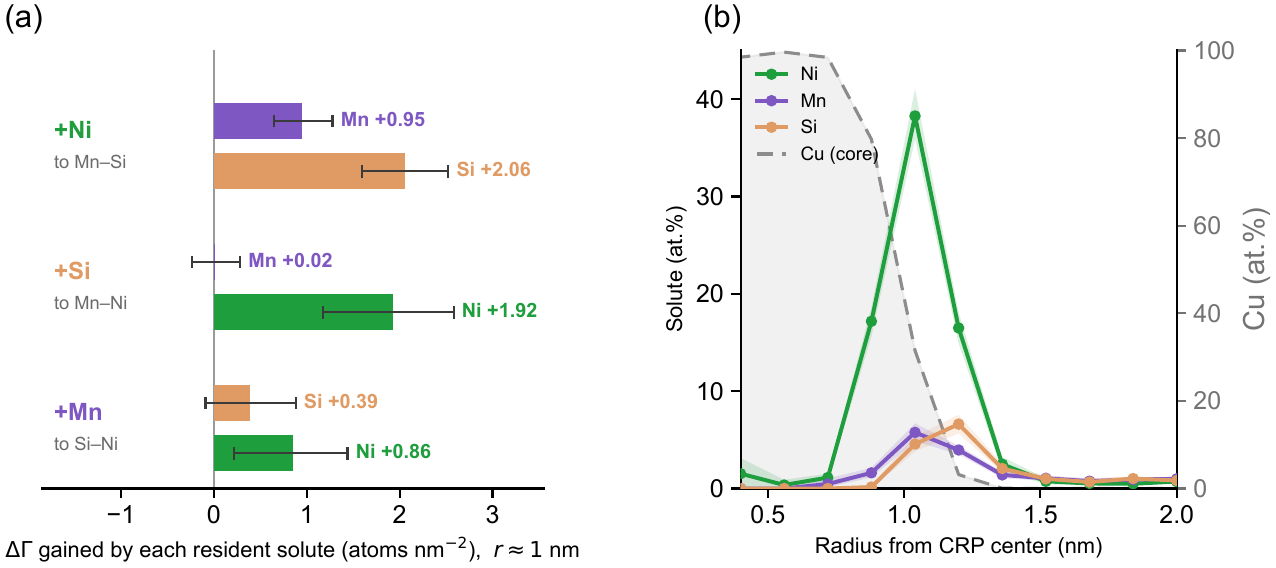}
\caption{Plots showing the composition dependence and location of co-segregation at the Fe/CRP interface at \SI{573}{\kelvin}. (a)~Change in interfacial excess \(\Delta\Gamma\) of each resident solute when the third solute is added to a binary condition, at a CRP radius \(r\approx\SI{1.0}{\nano\metre}\). (b)~Radial profiles of Ni, Mn and Si for the Mn--Ni--Si condition at the same radius, with the Cu profile on the right axis.}
\label{fig:cooperative}
\end{figure*}

The ternary enrichment relative to each binary state is isolated in \cref{fig:cooperative}. Adding Ni to Mn--Si increases both the Mn and Si excesses (by \SI{0.95}{} and \SI{2.06}{atoms.nm^{-2}} respectively), whereas adding Si to Mn--Ni or Mn to Ni--Si mainly increases the Ni excess (by \SI{1.92}{} and \SI{0.86}{atoms.nm^{-2}} respectively, \cref{fig:cooperative}a). The ternary response is therefore asymmetric and is not captured by adding independent solute contributions. The radial profiles further show that Ni, Mn and Si occupy the same shell at the CRP boundary instead of separate regions within the precipitate (\cref{fig:cooperative}b). Ni and Mn both peak at \(r=\SI{1.04}{\nano\metre}\), reaching \SI{38.3}{} and \SI{5.8}{at.\%} respectively, and Si peaks just beyond at \(r=\SI{1.20}{\nano\metre}\) with \SI{6.6}{at.\%}. All three peaks occur in the same narrow radial region at the CRP boundary.

\section{Discussion}
\label{sec:discussion}

\subsection{Interfacial order, co-segregation, and energy reduction}
\label{sec:discussion_order}

\begin{figure*}[ht]
\centering
\includegraphics[width=0.85\textwidth]{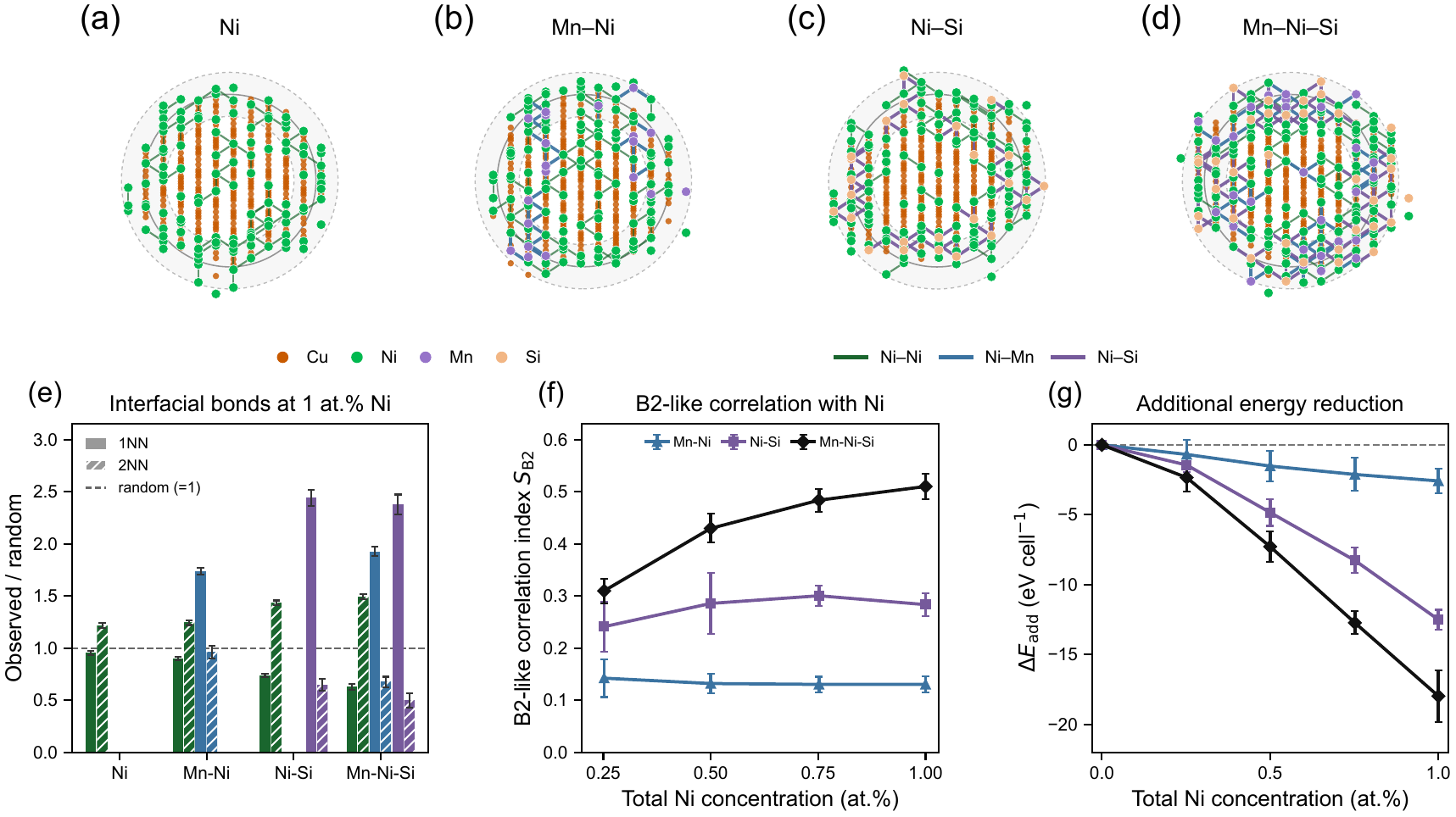}
\caption{Dependence of local order and energy on Ni at the Fe/CRP interface at \SI{573}{\kelvin}, with any Mn or Si held at 1 at.\%. (a--d)~Projected configurations at 1 at.\% total Ni for Ni, Mn--Ni, Ni--Si and Mn--Ni--Si. (e)~Observed interfacial bond populations divided by their random values at 1 at.\% Ni, at the first (1NN, solid) and second (2NN, hatched) neighbor shells. (f)~Local B2-like correlation index \(S_{\rm B2}\) for Mn--Ni, Ni--Si and Mn--Ni--Si, as total Ni increases from 0.25 to 1 at.\%. (g)~Paired formation-energy difference relative to the Ni-only system in \cref{eq:additional_energy}, as total Ni increases from 0 to 1 at.\%. Error bars give \(\pm1\) standard deviation across ten paired simulations. Further details are given in Supplementary Information, Section~S4.}
\label{fig:localorder}
\end{figure*}

In the previous section, we found that Ni promotes Mn and Si enrichment at the Fe/CRP boundary, but the segregation profiles show only where the solutes accumulate and not how they sit next to one another. To test whether local order at the boundary contributes to the stronger energy reduction when Mn and Si are present with Ni, we used calculations at one CRP size, holding Mn and Si at 1 at.\% and varying the total Ni content from 0 to 1 at.\%. The projected configurations at 1~at.\% Ni in \cref{fig:localorder}(a)--(d) show the Cu-rich core and the attached shell for Ni-only, Mn--Ni, Ni--Si and Mn--Ni--Si. The 1NN bonds between attached solutes use the same color scheme as the bond plot in \cref{fig:localorder}. In all four cases, Mn and Si atoms are mixed throughout the Ni-rich shell instead of forming separate regions or domains. The detailed Ni--Mn and Ni--Si bond statistics and their comparison with a random reference are given in \cref{fig:localorder}(e), and the local B2-like correlation index in (f).

The bond populations at 1~at.\% Ni quantify this arrangement (\cref{fig:localorder}e). Solid bars give the ratio of the observed number of a given 1NN bond type to the number expected from a random distribution of the same species over the same interfacial sites, and hatched bars give the same ratio at the 2NN shell. Ratios above one indicate enrichment and values below one indicate depletion. The Ni--Ni ratio is about \num{0.95} for the Ni-only case and decreases to about \num{0.90}, \num{0.74} and \num{0.63} for Mn--Ni, Ni--Si and Mn--Ni--Si, respectively. In contrast, Ni--Mn and Ni--Si bonds are enriched above random, with ratios of about \num{1.74} and \num{2.44} in the binaries and \num{1.93} and \num{2.38} in the ternary. The ternary interface therefore contains both types of unlike bonds and the fewest Ni--Ni contacts. At the 2NN shell the pattern reverses: the Ni--Ni ratio rises above one in every condition containing Mn or Si, while the Ni--Mn and Ni--Si ratios fall back to near random or below.

The local B2-like correlation index \(S_{\rm B2}\) quantifies this order across the attached shell (\cref{fig:localorder}f). At 1~at.\% Ni it is about \num{0.51} for Mn--Ni--Si and about \num{0.29} for Ni--Si, both well above the value near zero expected for random occupation, whereas Mn--Ni stays lower at about \num{0.13}. As the total Ni content increases from \num{0.25} to \num{1}~at.\%, \(S_{\rm B2}\) rises for Mn--Ni--Si but changes little for the two binary conditions, so increasing Ni strengthens B2-like order mainly in the ternary system. Because \(S_{\rm B2}\) stays well below \num{1}, the order is local and partial rather than a fully developed B2 phase.

The additional energy reduction \(\Delta E_{\rm add}\) becomes more negative as Ni increases in all three solute backgrounds (\cref{fig:localorder}g). At 1~at.\% Ni, it is about \SI{-2.6}{\electronvolt} per cell for Mn--Ni and \SI{-12.5}{\electronvolt} per cell for Ni--Si, compared with \SI{-18.0}{\electronvolt} per cell for Mn--Ni--Si. The ternary system combines stronger B2-like order with the largest additional energy reduction. The binary energies also decrease while their \(S_{\rm B2}\) values change little, so ordering is not the sole source of the energy reduction. Because \(\Delta E_{\rm add}\) includes energy changes throughout the complete simulation cell, it does not isolate an interfacial or ordering contribution. These bond and correlation statistics therefore describe the local arrangement within the solute population attached to the CRP, not coherent B2 order over the complete interface or an equilibrium B2 or G phase.

\subsection{Cu/MNSP energy, attachment, and shells}
\label{sec:discussion_wetting}

The clean Fe/CRP, Fe/MNSP and Cu/MNSP coefficients give the interface balance for attachment and partial coverage. The calculated coefficients give \(W_{\mathrm{adh},E}>0\), so replacing separate Fe/CRP and Fe/MNSP areas by Cu/MNSP contact lowers the internal-energy contribution. This attachment by itself does not ensure a concentric shell. Covering part of a CRP removes Fe/CRP area and creates both Cu/MNSP and Fe/MNSP area, so the balance is given by the spreading parameter in \cref{eq:spreading}. With the coherent bcc Cu representation of the CRP, complete coverage requires \(\gamma_{\rm Fe/CRP}\geq\gamma_{\rm Cu/MNSP}+\gamma_{\rm Fe/MNSP}\).

\begin{figure*}[ht]
  \centering
  \includegraphics[width=0.80\linewidth]{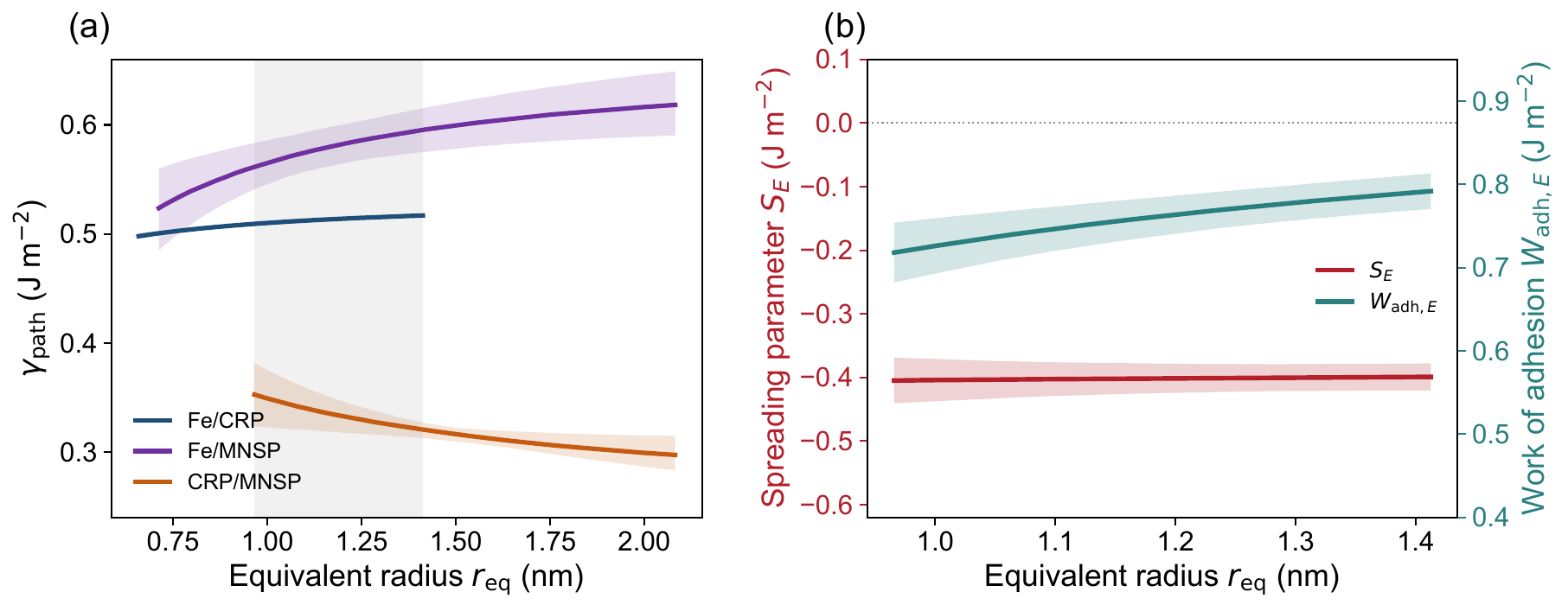}
\caption{Plots showing the wetting balance from coefficients based on internal energy over a common radius range. (a)~Coefficients \(\gamma_{\mathrm{path}}\) for Fe/CRP, Fe/MNSP and Cu/MNSP (labeled CRP/MNSP in the legend). The grey band marks the common radius range used in (b). (b)~Spreading parameter \({S_E}\) and work of adhesion \({W_{\mathrm{adh},E}}\) over that range. Fe/CRP is evaluated at \SI{573}{\kelvin} and the ordered phases at \SI{0}{\kelvin}.}
  \label{fig:equalradiusspreading}
\end{figure*}

Over the common size range, the three coefficients give \(S_E\approx\SI{-0.40}{\joule\per\square\metre}\) and \(W_{\mathrm{adh},E}=\SIrange{0.72}{0.79}{\joule\per\square\metre}\) (\cref{fig:equalradiusspreading}). The positive work of adhesion supports contact, whereas the negative spreading parameter penalizes complete coverage. This combination is consistent with a finite cap or appendage over the sampled size range \cite{smith1948grains,cahn1977critical}. The signs remain unchanged over the common radius range. This balance reflects only the internal-energy contribution defined in \cref{sec:methods_growth}.

Replacing the MNSP(G-phase) reference by NiMn B2 without vacancies keeps the signs of \(S_E\) and \(W_{\mathrm{adh},E}\) but moves \(S_E\) into the range \SIrange{-0.074}{-0.045}{\joule\per\square\metre}. The wetting balance is then much closer to complete coverage than for the MNSP(G-phase) result. This shift is consistent with the nearly closed shell in the trajectory without vacancies and the larger displacement from the center in the calculation using the MNSP(G-phase) composition (Supplementary Information, Fig.~S2). Because the two references also differ in Si content and sublattice chemistry, we do not try to separate the vacancy contribution here. The full B2 interface calculation is given in the Supplementary Information, Section~S10 and Fig.~S8.

The Gibbs adsorption result for Ni at the Fe/CRP boundary gives a consistent picture. As a sensitivity test, applying the Fe/CRP reduction from \cref{fig:isotherm} while keeping the Fe/MNSP and Cu/MNSP coefficients at their clean values shifts \(S_E\) and \(W_{\mathrm{adh},E}\) by the same amount. The spreading parameter becomes more negative and the work of adhesion remains positive, consistent with the clean-interface balance in \cref{fig:equalradiusspreading}. This comparison combines a free-energy change with clean internal-energy coefficients and is not a complete free-energy wetting calculation. The shift is smaller than the clean-interface value of \(S_E\approx\SI{-0.40}{\joule\per\square\metre}\), so a full free-energy balance would also include the Ni effects on Fe/MNSP and Cu/MNSP.

The total amount of Mn, Ni and Si relative to Cu, described by \(R_{\rm MNS/Cu}=(x_{\rm Mn}+x_{\rm Ni}+x_{\rm Si})/x_{\rm Cu}\), gives a second way to change the morphology. In the composition sweep reported in the Supplementary Information, Section~S6 and Figs.~S3 and S4, simulations starting from random, core--shell and appendage configurations all form a core--shell particle through \(R_{\rm MNS/Cu}=7\) and an appendage at \(R_{\rm MNS/Cu}=14\). The outcome follows the composition ratio and is insensitive to the initial geometry.

From the nominal compositions reported by Styman et al.\ for thermally aged RPV welds, \(R_{\rm MNS/Cu}\) is approximately \num{6.1} for the weld with lower Ni content and \num{8.6} for the weld with higher Ni content \cite{styman2015characterisation}. These values are closer to the core--shell condition at \(R_{\rm MNS/Cu}=7\) than to the appendage condition at \(R_{\rm MNS/Cu}=14\) in the controlled MC simulations. The comparison is consistent with their observation of centers enriched in Cu and interface regions enriched in Ni, Mn and Si. It remains qualitative because the nominal alloy ratio does not represent the composition of an individual precipitate.

\subsection{Comparison with prior models and experiment}
\label{sec:discussion_comparison}

These calculations build on earlier atomistic studies that examined Ni shells around Cu-rich precipitates and planar Fe/CRP interfaces. Curved-particle calculations with EAM and DFT potentials found Ni-enriched shells at the Fe/CRP boundary \cite{Seko2004,Gorbatov2015,Wang2019}, and planar DFT work examined how orientation, composition and co-segregation change the Fe/CRP energy \cite{Wang2020,Garrett2021,Garrett2022}. Those studies used relaxed \SI{0}{\kelvin} energy differences or prescribed interfacial occupations. The growth calculations here follow an equilibrated curved particle at finite temperature and allow all solutes to redistribute. Separate calculations at fixed Cu counts isolate the effect of Ni coverage, and the Fe--Ni Gibbs adsorption integral gives the relative change in Fe/CRP free energy at \SI{573}{\kelvin} (\cref{fig:isotherm}b,c and Supplementary Information, Section~S8 and Fig.~S6).

The Fe/CRP coefficient obtained here is close to the upper end of the Shu et al.\ nearest-neighbor broken-bond (NNBB) range for the same interface, whereas the Fe/MNSP and Cu/MNSP coefficients are much higher than their NNBB values \cite{shu2018thermodynamics} (\cref{tab:shucomparison}). \Cref{tab:shucomparison} also lists two checks that do not rely on \(\gamma_{\rm path}\), namely an independent flat-slab calculation of the same three interfaces (Supplementary Information, Sections~S2--S3) and \(\gamma_{\rm direct}\), evaluated from individual growth snapshots. For Fe/CRP, the flat-slab values and \(\gamma_{\rm direct}\) agree with \(\gamma_{\rm path}\) and with earlier DFT and kinetic Monte Carlo estimates for the same clean, coherent interface \cite{Wang2020,jourdan2010cluster}. For Fe/MNSP and Cu/MNSP, the flat-slab ranges fall below and above \(\gamma_{\rm path}\), while \(\gamma_{\rm direct}\) at large size is closer to the \(\gamma_{\rm path}\) values. Shu et al.\ used a disordered MnNiSi coating in their NNBB estimate and found \(S_{\rm MNSP}>0\), favoring complete wetting. In their own kinetic lattice Monte Carlo simulations, however, the coating stays disordered while a separate nucleus orders and grows into the appendage, which they interpreted as an ordered Mn--Ni--Si phase attached to a Cu-rich region. We find a similar separation in an unconstrained growth trajectory here, where the attached Mn, Ni and Si domain develops order similar to B2 as it grows. In the cluster-expansion calculation here, the ordered MNSP(G-phase) composition with structural vacancies is used at the interface and the full energy of a curved particle is evaluated, giving \(S_{\rm MNSP}<0\) and partial wetting. This approach therefore describes the Mn--Ni--Si boundary with more structural and chemical detail than a disordered coating, while remaining consistent with the appendage morphology found in the Shu et al.\ simulations.

The Gibbs adsorption result for Ni, the APT estimate of Isheim et al.\ \cite{Isheim2006}, and the Styman et al.\ measurements \cite{styman2015characterisation} all show that Ni segregation reduces the interface free energy. The Isheim et al.\ value of \SI{16}{\milli\joule\per\square\metre} is from a high-strength low-carbon steel. Styman et al.\ studied thermally aged RPV steel welds with bulk Ni contents of \SI{0.29}{at.\%} and \SI{1.66}{at.\%}, aged at \SIrange{603}{638}{\kelvin}, and report combined Ni--Mn--Si reductions of \SIrange{10}{17}{\milli\joule\per\square\metre} for the lower-Ni weld and \SIrange{53}{145}{\milli\joule\per\square\metre} for the higher-Ni weld \cite{styman2015characterisation}. Using the Gibbs integral (Supplementary Information, Fig.~S6c), the reduction for Ni only ranges from \SIrange{74}{150}{\milli\joule\per\square\metre} as the matrix content increases from \num{0.3} to \num{1}~at.\% Ni (nominal cell Ni content \SIrange{0.6}{1.5}{at.\%}), largely overlapping the range for the higher-Ni weld. The simulated Ni-enriched layer is about \SI{0.3}{\nano\metre} wide, between their measured interface widths of \SIrange{0.11}{0.13}{\nano\metre} and \SIrange{0.54}{0.56}{\nano\metre}.

 These experimental comparisons remain qualitative because the alloy compositions, temperatures, particle sizes and interface definitions differ from the simulations, and the reported reductions are segregation contributions, not absolute interface energies, so they do not by themselves determine \(S\) or \(W_{\rm adh}\). APT studies show CRP boundaries enriched in Ni in thermally aged welds \cite{styman2015characterisation}. Other APT work reports accelerated MNS precipitation in the presence of Cu and asymmetric CRP--MNSP particles \cite{wells2014evolution,Davis2020,almirall2020use}. The calculation here uses an integrated areal excess, which is less sensitive than a peak concentration to broadening during APT reconstruction \cite{marquis2010applications}. The combined observations are consistent with Ni accumulation at Fe/CRP boundaries and local environments enriched in Mn and Si.

\begin{table*}[t]
\centering
\caption{Comparison of the \(\gamma_{\rm path}\) coefficients with independent references, all in J\,m\(^{-2}\). The Shu et al.\ NNBB range is their nearest-neighbor broken-bond estimate for a disordered coating at \SI{0}{\kelvin} \cite{shu2018thermodynamics}. The flat-slab column gives static \SI{0}{\kelvin} coefficients (Supplementary Information, Sections~S2--S3), and the \(\gamma_{\rm direct}\) column gives values evaluated on individual growth snapshots.}
\label{tab:shucomparison}
\small
\resizebox{\ifdim\width>\linewidth\linewidth\else\width\fi}{!}{%
\begin{tabular}{lcccc}
\toprule
Interface & \(\gamma_{\rm path}\), finite particle & Shu et al.\ NNBB range, \SI{0}{\kelvin} & \(\gamma_{\rm flat}\) range, \SI{0}{\kelvin} & \(\gamma_{\rm direct}\) \\
\midrule
Fe/CRP   & 0.50--0.52, \SI{573}{\kelvin} & 0.317--0.518 & 0.45--0.55 & 0.519--0.522 \\
Cu/MNSP  & 0.30--0.36, \SI{0}{\kelvin}  & 0.076--0.124 & 0.35--0.53 & 0.318 \\
Fe/MNSP  & 0.52--0.62, \SI{0}{\kelvin}  & 0.081--0.133 & 0.49--0.56 & 0.579 \\
\bottomrule
\end{tabular}}
\end{table*}

\subsection{Physical sequence, thermodynamic scope, and limitations}
\label{sec:discussion_scope}
In this thermodynamic picture, the sequence starts with CRP precipitation. A small coherent CRP forms first and Ni then adsorbs at its boundary. As the Ni coverage increases, the interfacial energy decreases and Mn and Si also become enriched, with preferential Ni--Mn and Ni--Si bonds. In the simulated trajectory, this interfacial state is followed by an attached phase that grows into a compact asymmetric domain, while Ni remains on the exposed Fe/CRP interface. The binary calculations also show that the energy can decrease without a corresponding increase in \(S_{\rm B2}\). Local order accompanies this sequence but does not by itself account for the energy reduction.

This sequence does not imply that Ni physically transports Mn and Si or that a CRP must always form first. MNSPs can form at low Cu content in steels with high Ni after high irradiation fluence \cite{wells2014evolution,kamboj2026microstructure}, and irradiation defects provide other nucleation and transport pathways \cite{messina2020solute}. Our previous work showed that sufficient vacancy populations stabilize clusters rich in Mn, Ni and Si and change their ordering pathway \cite{Ke2025Vacancies,whiting2019understanding}. Pathways initiated by Cu precipitation and those associated with irradiation defects may therefore operate at the same time. Note that the sequence described here is a thermodynamic interpretation for one alloy and temperature, not a unique kinetic mechanism.

In this approach, we use a DFT-trained cluster expansion on a rigid bcc lattice. This Hamiltonian describes configurational interactions among multiple solutes but does not include lattice reconstruction, continuous strain, phonons or interfacial magnetic disorder. The analysis is therefore limited to coherent or nearly coherent bcc interfaces without large misfit stresses or diffuse Fe/Cu interphases. The ordered MNSP reference is the bcc mapping of the G-phase composition with structural vacancies. The NiMn B2 calculation gives the same signs of \(S_E\) and \(W_{\mathrm{adh},E}\) but changes their magnitudes, which shows that the wetting balance in this calculation depends on the local structure and composition of the MNSP boundary. Other possible Mn--Ni--Si interfacial states are beyond the scope of this work.

The interface coefficients and the resulting \(S_E\) and \(W_{\mathrm{adh},E}\) are based on excess internal energy. In addition, the Fe/MNSP and Cu/MNSP coefficients were evaluated at \SI{0}{\kelvin}, whereas the Fe/CRP coefficient was evaluated at \SI{573}{\kelvin}. This balance describes tendencies toward attachment and partial coverage. Determining an equilibrium contact angle requires all three interfacial free energies at the same temperature and diffusion potentials, together with the segregation contributions to the Fe/MNSP and Cu/MNSP interfaces. The Gibbs adsorption calculations include configurational entropy but give only relative Fe/CRP changes along separate Fe--Ni, Fe--Mn and Fe--Si paths, with one solute varied at a time. The present results therefore quantify how Ni changes the Fe/CRP free energy along this path and do not map out the full wetting behaviour over composition and temperature.

\subsection{Applications and outlook}
\label{sec:discussion_outlook}
Within this thermodynamic scope, the results connect segregation, interface free energy and precipitate morphology in this alloy system. At grain boundaries, solute enrichment is used to adjust chemistry, structure and cohesion \cite{raabe2014segregation,kirchheim2002grain}. Similar ideas apply at precipitate interfaces, where curvature and changing composition add further complexity. Here, we use one Hamiltonian to calculate the three interface coefficients and their dependence on particle size, and the Gibbs adsorption relation separately provides the relative change in Fe/CRP free energy caused by Ni \cite{gibbs1878equilibrium,cahn1979thermodynamics,kaplan2013review,kirchheim2007reducing}. These quantities give a consistent picture of how segregation and interface energies affect precipitation models and the capillarity driving force for coarsening.

The calculated coefficients serve as interface-energy parameters for precipitation models. In high-strength steels, Cu--Ni co-precipitation refines and stabilizes the strengthening dispersion \cite{jiao2013synergistic}. In irradiated low-alloy steels, Cu-rich clusters promote the Mn--Ni--Si phases that control hardening at later stages \cite{wells2014evolution,mamivand2019cumnnisi}. The dependence of \(\gamma_{\rm path}\) on particle size therefore provides an internal-energy-based coefficient for cluster dynamics, nucleation and phase field models \cite{ke2017thermodynamic,deschamps2021precipitation,zhao2022coprecipitated}. For Fe/CRP, the Gibbs adsorption result provides the relative change with Ni. Using atomistic coefficients in place of adjustable values can reduce uncertainty in predictions of strengthening and embrittlement in low-alloy steels and support microstructure engineering and life-assessment calculations that connect alloy chemistry, precipitate populations and mechanical response.

This work focuses on low-alloy steels, but similar questions also arise in more concentrated alloys. Concentrated and high-entropy alloys are being considered for reactor components because of their potential radiation tolerance \cite{zinkle2013materials,george2019high,zhang2015disorder}. Segregation at their grain boundaries has been related to the local boundary energy \cite{li2020grain}, whereas this relation is less well characterized at precipitate interfaces. Radiation-induced segregation can also drive interfaces away from equilibrium in structural materials, changing the local composition at defects and solute clusters \cite{messina2020solute,ke2023microstructure}. Future work could combine the dependence on diffusion potential with the contribution from solute fluxes and examine how both change interface energy and morphology.

\section{Conclusions}
\label{sec:conclusions}

In this work we quantified how selective Ni segregation and Mn--Ni--Si co-segregation at coherent Fe/Cu interfaces change interface energies. We then related these changes to the shell-to-appendage morphology of Cu-rich and Mn--Ni--Si-rich precipitates in low-alloy steels. A multicomponent cluster expansion combined with lattice Monte Carlo simulations was used to obtain Fe/CRP, Fe/MNSP and Cu/MNSP interface coefficients from one Hamiltonian. These coefficients based on internal energy are about \(0.50\)--\(0.52\)~J~m\(^{-2}\) for Fe/CRP, \(0.52\)--\(0.62\)~J~m\(^{-2}\) for Fe/MNSP and \(0.30\)--\(0.36\)~J~m\(^{-2}\) for Cu/MNSP. The balance \(\gamma_{\rm Cu/MNSP}<\gamma_{\rm Fe/CRP}\lesssim\gamma_{\rm Fe/MNSP}\), together with a positive work of adhesion and a negative spreading parameter, is consistent with an attached Mn--Ni--Si domain that partially covers a Cu-rich precipitate.

The simulations show that Ni develops a strong interfacial excess at the Fe/CRP boundary, forming an approximately monolayer-thick adsorption zone (\SIrange{7.4}{9.6}{atoms.nm^{-2}} in the growth calculations). Mn and Si enrich the same region when Ni is present, and the attached shell develops local B2-like order that combines Ni--Mn and Ni--Si unlike bonds. The Gibbs adsorption analysis for Fe/CRP indicates that this segregation lowers the interface free energy by about \(\SI{0.15}{\joule\per\square\metre}\) at a matrix Ni content of \(\SI{1}{at.\%}\), in broad agreement with experimental estimates for high-Ni steels. Growth calculations that vary the Mn, Ni and Si contents relative to Cu connect this balance to morphology: lower Mn--Ni--Si contents relative to Cu favor a Ni- and Mn-enriched shell around the Cu-rich core, whereas higher contents favor an attached Mn--Ni--Si precipitate on the CRP.

Within the internal-energy and mixed-temperature scope of this work, these results give a thermodynamic picture of how segregation, local order and interface energies are linked to the core--shell and appendage morphologies observed in low-alloy steels. The additional internal-energy reduction from co-segregation increases in the order Mn--Ni \ensuremath{<} Ni--Si \ensuremath{<} Mn--Ni--Si, while P develops only a small excess with little further change in energy. The interface coefficients and segregation trends obtained here can be incorporated into precipitation and embrittlement models to relate alloy chemistry and irradiation history to the evolution of Cu--Mn--Ni--Si co-precipitates and their influence on mechanical response. Segregation-induced changes in interface energies can then be used to reduce the capillarity driving force for coarsening and help stabilize a fine precipitate dispersion.

\section*{CRediT authorship contribution statement}

\textbf{Jia-Hong Ke.} Writing -- review \& editing, Writing -- original draft, Visualization, Software, Project administration, Methodology, Investigation, Funding acquisition, Formal analysis, Data curation, Conceptualization.
\textbf{Anshul Kamboj.} Writing -- review \& editing, Visualization, Investigation, Formal analysis.
\textbf{Mukesh Bachhav.} Writing -- review \& editing, Resources, Project administration, Investigation, Funding acquisition, Conceptualization.

\section*{Declaration of competing interest}

The authors declare that they have no known competing financial interests or personal relationships that could have appeared to influence the work reported in this paper.

\section*{Data availability}

The data that support the findings of this study are available from the corresponding author upon reasonable request.

\section*{Acknowledgements}

This study was funded by the Laboratory Directed Research and Development (LDRD) Program at Idaho National Laboratory under grant number 23A1070-092FP. This research was also supported by the Nuclear Science User Facilities Rapid Turnaround Experiment (NSUF RTE) under NSUF 24-4923. This manuscript has been authored by Battelle Energy Alliance, LLC under Contract No. DE-AC07-05ID14517 with the U.S. Department of Energy. The authors would like to thank Prof. G. Robert Odette for many helpful discussions. The United States Government retains, and the publisher, by accepting the article for publication, acknowledges that the United States Government retains a nonexclusive, paid-up, irrevocable, world-wide license to publish or reproduce the published form of this manuscript, or allow others to do so, for United States Government purposes. This research made use of the resources of the High Performance Computing Center at Idaho National Laboratory, which is supported by the Office of Nuclear Energy of the U.S. Department of Energy and the Nuclear Science User Facilities under Contract No. DE-AC07-05ID14517.

\bibliographystyle{elsarticle-num}
\bibliography{references}

\clearpage
\onecolumn
\appendix
\begingroup
\renewcommand{\thesection}{S\arabic{section}}\renewcommand{\thefigure}{S\arabic{figure}}
\renewcommand{\thetable}{S\arabic{table}}\renewcommand{\theequation}{S\arabic{equation}}
\setcounter{section}{0}\setcounter{figure}{0}\setcounter{table}{0}\setcounter{equation}{0}
\makeatletter\renewcommand{\@biblabel}[1]{[S#1]}\makeatother
\begin{center}
{\Large\bfseries Supplementary Information for\\ Interplay of interfacial energetics and selective Ni segregation in the morphology of co-precipitates in low-alloy steels\par}\vskip 1.5em
{\large Jia-Hong Ke$^{a}$\quad Anshul Kamboj$^{b}$\quad Mukesh Bachhav$^{b}$\\[6pt]
\begin{minipage}{0.9\textwidth}\centering\small
$^{a}$Computational Mechanics and Materials Department, Idaho National Laboratory, Idaho Falls, ID 83415, USA\\
$^{b}$Nuclear Materials Performance Department, Idaho National Laboratory, Idaho Falls, ID 83415, USA\\[3pt]
Corresponding author \href{mailto:jiahong.ke@inl.gov}{jiahong.ke@inl.gov}
\end{minipage}\par}
\end{center}
\vskip 1em

This Supplementary Information reports numerical criteria, uncertainty estimates and supporting calculations used in the main text. CRP denotes a Cu-rich precipitate and MNSP a Mn--Ni--Si-rich precipitate, as in the main text. Section~\ref{si:methods} summarizes additional methods and numerical checks. Sections~\ref{si:mnspflatslab}--\ref{si:mnspcrosscheck} give flat-slab interface coefficients and their comparison with earlier work, Section~\ref{si:localorder} describes the local-order analysis, Sections~\ref{si:morphvacancy} and~\ref{si:seeded} examine the effects of structural vacancies and the amount of Mn, Ni and Si relative to Cu, and Sections~\ref{si:mclean}--\ref{si:b2wetting} support the segregation analysis and wetting balance.

\section{Additional methods and numerical checks}
\label{si:methods}

The cluster expansion (CE) and Monte Carlo (MC) methods are described in Section~2.1 of the main text. Here we summarize the additional criteria used in the growth calculations. Clusters were identified using the bcc first-nearest-neighbor (1NN) distance of \SI{0.255}{\nano\metre}. A configuration was kept when the precipitate core contained at least \num{100} sites and no second connected component contained ten or more atoms. For each reported particle size, we used accepted states from at least eight of the ten independent simulations. The ordered MNSP calculations used the same criteria and ten independent simulations. In the CE database, only relaxed structures that remained close to the parent bcc lattice were kept. Structures with a lattice deformation greater than \num{0.1}\% or that reconstructed away from bcc were excluded.

Using the growth path fit defined in Section~2.2 of the main text, we first averaged the energies at each controlled size so that every size had equal weight. We then resampled the ten complete growth trajectories \num{2000} times to estimate the sampling uncertainty. Resampling complete trajectories preserves the relation among particle sizes within each simulation.

The clean Fe/CRP calculation at \SI{573}{\kelvin} contained no Ni or other solutes. The selected fit gives \(\gamma_{\rm path}=\SIrange{0.498}{0.517}{\joule\per\square\metre}\) over \(r_{\rm eq}=\SIrange{0.659}{1.413}{\nano\metre}\). This corresponds to the rounded range of \SIrange{0.50}{0.52}{\joule\per\square\metre} in the main text. We also compared several ways of treating the particle size using the same reference energies. A fit containing an area term and a constant offset gives \SI{0.512}{\joule\per\square\metre}. Adding the first inverse radius contribution in the integrated Tolman form \cite{SI:Tolman1949} gives \SI{0.533}{\joule\per\square\metre}. Expanding \(E_{\rm xs}/A\) through second order in inverse radius gives \SIrange{0.533}{0.538}{\joule\per\square\metre} depending on the area weighting. Direct averages of \(E_{\rm xs}/A\) give \SIrange{0.519}{0.522}{\joule\per\square\metre}. The \SIrange{0.512}{0.538}{\joule\per\square\metre} values summarize how the fitted coefficient depends on the treatment of particle size and do not represent a statistical confidence interval.

The same fit was used for the ordered Fe/MNSP calculation at \SI{0}{\kelvin}. After applying the acceptance criteria, 97 particle sizes from the ten simulations were included. These sizes range from 128 to 3200 precipitate sites, with \(r_{\rm eq}\) between \SI{0.712}{} and \SI{2.082}{\nano\metre}. The resulting \(\gamma_{\rm path}\) ranges from \SI{0.524}{} to \SI{0.619}{\joule\per\square\metre}. This corresponds to the rounded range of \SIrange{0.52}{0.62}{\joule\per\square\metre} reported in the main text.

The same capillarity form was used for the Cu/MNSP calculation at \SI{0}{\kelvin}. After applying the acceptance criteria, the growth path gives \(\gamma_{\rm path}\) between \SI{0.30}{} and \SI{0.36}{\joule\per\square\metre} over the accepted sizes. This is the rounded range reported in the main text. We tested the same alternative treatments of particle size as for Fe/MNSP and obtained shifts comparable to those in the Fe/MNSP case.

To see how Ni coverage changes with particle size, we simulated \num{55} conditions at eleven nominal Ni concentrations from 0 to 1~at.\% and five Cu counts. The Cu counts corresponded to \(r_{\rm eq}=\SI{0.69}{}, \SI{0.80}{}, \SI{1.00}{}, \SI{1.19}{}\) and \SI{1.39}{\nano\metre}. Ten independent simulations were performed for each condition, giving \num{550} simulations in total. At a fixed Cu count, the radius varied by no more than \num{0.40}\% across the Ni concentrations. At each Cu count, \(\Delta\bar g_{\rm Ni}\) was defined as the mean formation-energy change relative to the no-Ni calculation at the same Cu count, divided by \(A_{\rm eq}\). It is an internal-energy difference and not an absolute interfacial free energy.

 We used homogeneous Fe--Ni semigrand-canonical calculations to relate the matrix composition to the Ni--Fe diffusion potential. Four forward and four reverse scans used a \(20\times20\times20\) cell. At the two endpoint compositions, a separate \(30\times30\times30\) calculation differed by no more than \SI{0.0014}{\electronvolt}, well below the \SI{0.01}{\electronvolt} tolerance. The Gibbs adsorption uncertainty was obtained by jointly resampling the interfacial excess calculations and the semigrand-canonical scans. In each resample, the endpoints were remapped and only paths that covered the measured composition range were kept for the uncertainty estimate.

The calculations over the wider composition range reported in Section~3.3 used separate Gibbs paths for Ni, Mn and Si. The homogeneous SGC scans covered the diffusion potential range containing the target dilute composition and used the same multicomponent CE, \(20\times20\times20\) cell and four forward and four reverse scans. For each solute, the interfacial excess was measured in canonical alloying calculations using a single Cu-rich precipitate with \(r_{\rm eq}\approx\SI{1.0}{\nano\metre}\) embedded in Fe. The matrix content of one solute at a time was increased by replacing Fe. The complete Gibbs paths and integration are reported in Section~S8.

For the morphology analysis, precipitates in the large cell were identified using a first-nearest-neighbor (1NN) cutoff of \SI{0.25}{\nano\metre} and a minimum of \num{80} atoms. After exact periodic tiles were collapsed, we tracked the dominant Cu component backward from the overlap among its sites. We used atomic Wigner--Seitz face areas to divide the CRP boundary into regions contacting the MNSP, enriched in solutes or rich in Fe. The Cu displacement was normalized by the radius of the attached MNSP. Configurations between \num{20} and \num{5000} MC passes were analyzed and rendered with OVITO \cite{SI:Stukowski2010OVITO}.

The definitions and thermodynamic scope of the interfacial quantities are summarized in Table~\ref{tab:quantities}. The quantities obtained from the growth calculations are based on excess internal energy, whereas the Gibbs adsorption analysis gives a relative change in interfacial free energy between two matrix states. Together these quantities represent the internal-energy and free-energy parts of the interface description used in the main text. Here \(E_{\rm xs}\) is the excess internal energy, \(N_{\rm C}\) is the number of precipitate sites, \(r_{\rm eq}\) is the equivalent-sphere radius defined from \(N_{\rm C}\) and the bcc site density, and \(A=4\pi r_{\rm eq}^2\) is the corresponding area. The spreading parameter \(S_E\) and work of adhesion \mbox{\(W_{\mathrm{adh},E}\)} use the same expressions as in the main text and are evaluated over the common radius range.

\begingroup
\section{Flat-slab interface coefficients}
\label{si:mnspflatslab}

\begin{table}[t]
\centering
\caption{Definitions and thermodynamic scope of the interfacial quantities used in the main text.}
\label{tab:quantities}
\footnotesize\linespread{1}\selectfont
\begin{tabular}{p{0.22\textwidth}p{0.28\textwidth}p{0.40\textwidth}}
\toprule
Quantity & Definition & Interpretation and scope \\
\midrule
\(\gamma_{\rm path}(r)=\mathrm{d}E_{\rm xs}/\mathrm{d}A\) & Local slope of the capillarity fit to \(E_{\rm xs}(A)\) & Coefficient based on internal energy and evaluated as a function of particle size. It is insensitive to a constant energy offset and is not an absolute interfacial free energy \\
\(\gamma_{\rm direct}\equiv E_{\rm xs}/A\) & Excess energy relative to the reference states divided by the area of the equivalent sphere & Average coefficient for the entire particle based on excess internal energy and including any constant offset \\
\(\gamma_{\rm flat}=(E_{\rm tot}-N_{\rm m}\mu_{\rm m}-N_{\rm p}g_{\rm p})/(2A)\), \cref{eq:si_flat_gamma} & Energy of a periodic flat-slab bicrystal relative to the same matrix and precipitate references, per unit interface area & Static \SI{0}{\kelvin} comparison that does not use the growth trajectory, evaluated per crystallographic orientation and termination pair (Section~S3) \\
\(\Delta\bar g_{\rm Ni}\) & Canonical internal-energy change per equivalent area relative to the state without Ni at the same Cu count & Adsorption trend based on internal energy at \SI{573}{\kelvin}. It separates coverage from size and does not include configurational entropy \\
\(\Delta\gamma_{\rm Fe/CRP}\) & \(-\int\Gamma_i^{(\rm Cu)}\,\mathrm{d}\Delta\mu_i\) & Relative change in Fe/CRP interfacial free energy along a specified single-solute matrix-composition change \\
\(\Delta G_{\rm seg}=-k_{\rm B}T\ln K\) & Langmuir--McLean fit to the \SI{573}{\kelvin} isotherm at \(r_{\rm eq}=\SI{1.0}{\nano\metre}\) & Segregation free energy from a Langmuir--McLean fit at \SI{573}{\kelvin} \\
\(\Delta E_{\rm add}^{X}\) & Paired difference between formation-energy changes for \(X\)+Ni and Ni only & Formation-energy comparison for the complete simulation cell, including energy changes throughout the cell \\
\bottomrule
\end{tabular}
\end{table}
As a separate check on the interface coefficients, we computed static \SI{0}{\kelvin} flat interfaces using the same cluster-expansion energy model. Flat slabs were built for Fe/CRP, Fe/MNSP and Cu/MNSP.

Each flat-slab cell is a fully periodic bicrystal. One block contains disordered bcc Fe or Cu, and the other contains either Cu or ordered MNSP (Cu for Fe/CRP, MNSP for Fe/MNSP and Cu/MNSP). The blocks are stacked along one crystallographic direction, and separate cells were constructed for the \{100\}, \{110\} and \{111\} orientations. Because the cell repeats along the stacking direction, it contains two matrix/precipitate boundaries related by translation, so the interfacial coefficient for each slab is
\begin{equation}
\gamma_{\rm flat}=\frac{E_{\rm tot}-N_{\rm m}\mu_{\rm m}-N_{\rm p}g_{\rm p}}{2A},
\label{eq:si_flat_gamma}
\end{equation}
where \(E_{\rm tot}\) is the total energy of the periodic cell, \(N_{\rm m}\) and \(N_{\rm p}\) are the site counts on the matrix and precipitate sides, \(\mu_{\rm m}\) and \(g_{\rm p}\) are the same matrix and precipitate reference energies per site used throughout this work (\cref{si:methods}), and \(A\) is the cross-sectional area of the periodic cell.

When the two periodic boundaries have inequivalent terminations, \(\gamma_{\rm flat}\) is their arithmetic mean and does not represent either termination separately. We therefore identify each MNSP calculation by its termination pair.

For Fe/CRP, both blocks are a single chemical species. Each orientation then has two equivalent boundaries and no choice of termination, so three calculations, one per orientation, are sufficient. Their values are used directly in Section~\ref{si:mnspcrosscheck}.

For the MNSP slabs, each block was built so that the Ni\(_{16}\)Mn\(_6\)Si\(_7\)Va\(_3\) composition is exact. The repeat period along the stacking direction is eight layers for \{100\} and \{111\} and two layers for \{110\}. We used one MNSP period for \{100\} and \{111\} and two periods (four layers) for \{110\} so that all three orientations have a similar physical thickness. The Fe and Cu blocks contain ten layers for \{100\} and \{111\} and five layers for \{110\}, which keeps the total cell sizes comparable across orientations.

Because the eight-layer MNSP period contains several compositionally distinct layers, the layer next to the matrix depends on where the repeat is cut. For each orientation we therefore built every geometrically and compositionally distinct termination pair, for example Mn/Ni, Ni/Mn and, for \{100\}, Si/Ni boundaries. A few pairs are related by the inversion symmetry that exchanges the two interfaces in one periodic cell and give the same \(\gamma_{\rm flat}\). These are counted once when we quote ranges.  \Cref{fig:mnspflatslabsnapshots} shows snapshots of all nine constructed termination pairs. Three rows (\{100\} Si, \{111\} Mn and \{111\} Si, vacancy-mixed) pair with a second interface that is chemically pure Ni at the surface but differ from the Ni-labeled rows because the layers beneath the interface are different.

We report every MNSP flat-slab calculation individually, each paired with an atomic configuration of the interface. Section~\ref{si:mnspcrosscheck} summarises these flat-interface results.
\endgroup

\begingroup
\section{Flat-slab interface energies and literature comparison}
\label{si:mnspcrosscheck}
For Fe/CRP, the static \SI{0}{\kelvin} flat-slab calculation of \cref{si:mnspflatslab} gives coefficients of \SI{0.549}{}, \SI{0.446}{} and \SI{0.540}{\joule\per\square\metre} for \{100\}, \{110\} and \{111\}. \Cref{tab:fecrporientation} compares these values with earlier calculations for the same clean, coherent bcc-Fe/Cu interface. Most of those studies used density-functional theory or an embedded-atom potential. Shu et al.\ reported nearest-neighbor broken-bond estimates, and Jourdan et al.\ gave a 0~K broken-bond value from atomistic kinetic Monte Carlo. In every source that resolves more than one orientation, \{110\} has the lowest energy among the reported facets. Our \{110\} coefficient of \SI{0.446}{\joule\per\square\metre} is close to the DFT value of about \SI{0.45}{\joule\per\square\metre} from Wang et al.\ and remains within the range of literature values in \cref{tab:fecrporientation}.

\begin{table}[t]
\centering

\caption{Static \SI{0}{\kelvin} Fe/CRP interfacial energies from flat slabs and earlier calculations, all in \si{\joule\per\square\metre}.}
\label{tab:fecrporientation}
\begin{tabular}{lccc}
\toprule
Source & \{100\} & \{110\} & \{111\} \\
\midrule
This work & 0.549 & 0.446 & 0.540 \\
Wang et al.\ \cite{SI:Wang2020} (DFT) & \(\approx\)0.61 & \(\approx\)0.45 & \(\approx\)0.52 \\
Wang et al.\ \cite{SI:Wang2020CompMatSci} (DFT) & 0.54 & 0.39 & 0.44 \\
Garrett and Race \cite{SI:GarrettRace2021} (DFT) & 0.530 & 0.350 & 0.430 \\
Seko et al.\ \cite{SI:Seko2004Calphad} (DFT) & 0.60 & 0.24 & 0.34 \\
Shu et al.\ \cite{SI:shu2018thermodynamics} (NNBB) & 0.449 & 0.317 & 0.518 \\
Jourdan et al.\ \cite{SI:jourdan2010cluster} (kMC) & -- & 0.431 & -- \\
Ludwig et al.\ \cite{SI:Ludwig1998} (EAM) & 0.318 & 0.121 & -- \\
\bottomrule
\end{tabular}
\end{table}

Flat-slab calculations for MNSP give a corresponding set of interface energies. Across the six distinct MNSP calculations, \(\gamma_{\rm flat}\) covers \SI{0.49}{} to \SI{0.56}{\joule\per\square\metre} for Fe/MNSP and \SI{0.35}{} to \SI{0.53}{\joule\per\square\metre} for Cu/MNSP (\cref{tab:mnspflatslab}). Among these, the most termination-sensitive case is Cu/MNSP \{111\} with a Ni/Si-terminated boundary (\cref{fig:mnspflatslabsnapshots}p), at \SI{0.527}{\joule\per\square\metre}. This value is closer to the Fe/MNSP range than to the other two Cu/MNSP \{111\} terminations (\SI{0.35}{} and \SI{0.43}{\joule\per\square\metre}).

\begin{table}[htbp]
\centering

\caption{Static \SI{0}{\kelvin} flat-slab coefficients for the nine MNSP termination pairs, in \si{\joule\per\square\metre}, paired with the interface snapshots in \cref{fig:mnspflatslabsnapshots}. Termination 1 is the composition each row is labeled for and is built against the Fe or Cu block. Termination 2 is the composition at the cell's other boundary. Each coefficient is the arithmetic mean for the two periodic boundaries. Rows marked * share the same \(\gamma_{\rm flat}\) as the row above.}
\label{tab:mnspflatslab}
\small
\resizebox{\ifdim\width>\linewidth\linewidth\else\width\fi}{!}{%
\begin{tabular}{lllcccc}
\toprule
Orientation & Termination 1 & Termination 2 & Fe/MNSP & Snapshot & Cu/MNSP & Snapshot \\
\midrule
\{100\} & Mn & Ni & 0.491 & (a) & 0.368 & (j) \\
\{100\} & Ni* & Mn & 0.491 & (b) & 0.368 & (k) \\
\{100\} & Si & Ni & 0.545 & (c) & 0.402 & (l) \\
\{110\} & vacancy-mixed & dense & 0.509 & (d) & 0.347 & (m) \\
\{110\} & dense* & vacancy-mixed & 0.509 & (e) & 0.347 & (n) \\
\{111\} & Mn & Ni & 0.551 & (f) & 0.346 & (o) \\
\{111\} & Ni & Si, dense & 0.548 & (g) & 0.527 & (p) \\
\{111\} & Si, dense* & Ni & 0.548 & (h) & 0.527 & (q) \\
\{111\} & Si, vacancy-mixed & Ni & 0.557 & (i) & 0.429 & (r) \\
\bottomrule
\end{tabular}}
\end{table}

\begin{figure}[htbp]
\centering

\includegraphics[width=\textwidth]{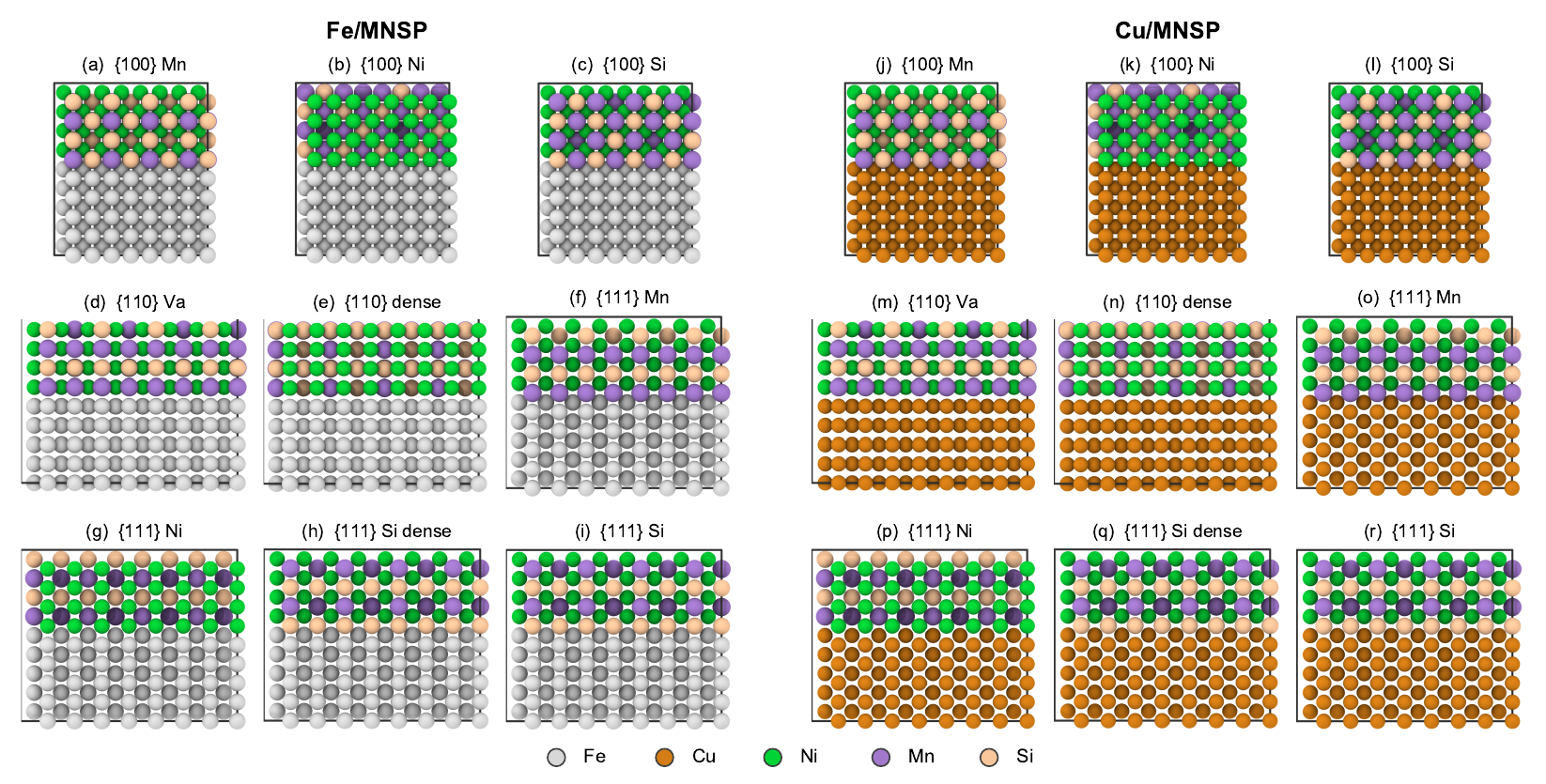}
\caption{Atomic configurations of the interface for the nine MNSP flat-slab termination pairs. Snapshots are arranged as two \(3\times3\) blocks (Fe/MNSP left, Cu/MNSP right), one image per row of \cref{tab:mnspflatslab}. Colors follow the key at the bottom (Fe grey, Cu orange, Ni green, Mn lavender, Si peach).}
\label{fig:mnspflatslabsnapshots}
\end{figure}

Shu et al.\ also report nearest-neighbor broken-bond (NNBB) estimates for the Fe/MnNiSi and Cu/MnNiSi interfaces at the same three orientations. Their model uses a disordered MnNiSi coating, whereas this study uses an ordered, vacancy-containing MNSP structure \cite{SI:shu2018thermodynamics}. \Cref{tab:mnspshuorientation} places their values next to the flat-slab range of \cref{tab:mnspflatslab}. Every flat-slab value is higher than the corresponding NNBB estimate, by factors of a few for all three orientations and both hosts.

\begin{table}[t]
\centering

\caption{Fe/MNSP and Cu/MNSP flat-slab ranges by orientation, from \cref{tab:mnspflatslab}, compared with the nearest-neighbor broken-bond (NNBB) estimates of Shu et al.\ \cite{SI:shu2018thermodynamics}, all in \si{\joule\per\square\metre}.}
\label{tab:mnspshuorientation}
\resizebox{\ifdim\width>\linewidth\linewidth\else\width\fi}{!}{%
\begin{tabular}{lcccc}
\toprule
& \multicolumn{2}{c}{Fe/MNSP} & \multicolumn{2}{c}{Cu/MNSP} \\
Orientation & Flat-slab range & Shu et al.\ NNBB & Flat-slab range & Shu et al.\ NNBB \\
\midrule
\{100\} & 0.491--0.545 & 0.115 & 0.368--0.402 & 0.107 \\
\{110\} & 0.509 & 0.081 & 0.347 & 0.076 \\
\{111\} & 0.548--0.557 & 0.133 & 0.346--0.527 & 0.124 \\
\bottomrule
\end{tabular}}
\end{table}

\endgroup

\section{Local order and additional energy reduction at the Fe/CRP interface}
\label{si:localorder}

We examined how Ni changes local order at a fixed CRP size using the calculations described in Section~2.5 of the main text at \(r_{\rm eq}\approx\SI{1.0}{\nano\metre}\). Each \(30\times30\times30\) bcc cell contained \num{27000} lattice sites and a Cu precipitate with 354 atoms, which corresponds to this radius from the equivalent-sphere relation in the main text. The total Ni concentration was varied from 0 to 1~at.\% in five steps, while Mn and Si were each kept at 1~at.\% when included. For each composition and chemistry, we performed ten independent simulations. Simulations with the same ensemble index used the same Cu sites, and the Ni sites were chosen from a shared list to allow direct comparisons. Every condition was equilibrated separately, and no trajectory was continued from a lower Ni concentration. No ordered interfacial layer was imposed. After \num{5000} equilibration passes, we sampled 20 configurations at intervals of 250 passes during a further \num{5000} passes.

Ni, Mn and Si atoms were grouped into components connected through 1NN bonds, and a component was counted as attached only when at least one atom had direct 1NN contact with the strict Cu core. Only bonds within this attached population were included.

The observed bond counts were compared with a random reference for the Ni-only, Mn--Ni, Ni--Si and Mn--Ni--Si systems. This analysis uses a fixed set of interface sites containing every bcc site outside the strict Cu core that has direct 1NN contact with the core, with both Fe and solute sites included. For the random reference, the sites, the bonds between them and the number \(n_i\) of each species were held fixed, while the species were distributed randomly over the \(N\) eligible sites. For \(N_{\rm e}\) bonds, the expected number of \(i\)--\(i\) pairs is \(N_{\rm e}n_i(n_i-1)/[N(N-1)]\) and the expected number of \(i\)--\(j\) pairs is \(2N_{\rm e}n_i n_j/[N(N-1)]\). This construction compares the Ni--Ni population in the Ni-only system with random occupancy of the same interface sites and bonds. The ratios shown in the local order figure of the main text were averaged over the 20 sampled configurations from each paired simulation at 1~at.\% total Ni. All 1NN counts used a \SI{2.55}{\angstrom} cutoff.

To distinguish B2-like order from a general preference for unlike neighbors, we counted both first-neighbor (1NN) and second-neighbor (2NN) bonds. In a bcc lattice, 1NN bonds connect opposite sublattices, whereas 2NN bonds connect sites on the same sublattice. B2-like order therefore gives more Ni--Mn and Ni--Si pairs at 1NN and more like pairs, including Ni--Ni, at 2NN. The 2NN shell covered distances from \SI{2.60}{} to \SI{3.00}{\angstrom} and used the same fixed interface sites as the 1NN analysis. This reversal between the two shells is commonly used to identify B2 order in bcc alloys \cite{SI:stana2016,SI:kormann2021,SI:xu2023,SI:woodgate2025}.

For neighbor shell \(m\), \(f_{\rm unlike}^{m}\) is the fraction of bonds joining Ni to Mn or Si, and \(c_m\) is the fraction of Ni atoms at the two ends of these bonds, with both ends counted. The shell correlation is
\begin{equation}
\rho_m=1-\frac{f_{\rm unlike}^{m}}{2c_m(1-c_m)}.
\label{eq:rho_shell}
\end{equation}
This Warren--Cowley-type correlation compares the observed fraction of unlike bonds with the value expected for random mixing at the same shell composition \cite{SI:cowley1950,SI:defontaine1994}. A negative value means that unlike bonds are favored, while a positive value means that like bonds are favored. The correlation can be calculated only when Ni and at least one of Mn or Si are present. It is therefore not reported at zero total Ni.

We combine the two shells as \(S_{\rm B2}=\tfrac{1}{2}(\rho_{\rm 2NN}-\rho_{\rm 1NN})\). Random occupation gives \(S_{\rm B2}\approx0\), while an equiatomic ideal B2 domain gives \(S_{\rm B2}=1\). At other compositions, the possible range depends on composition and on the bonds present at the finite interface. We therefore use \(S_{\rm B2}\) to measure the 1NN--2NN reversal relative to random occupation.

To check the random reference, we randomly reassigned the species among the same attached sites while keeping the number of each species fixed. This was repeated for all 20 sampled configurations from each of the ten independent simulations at 1~at.\% Ni. For the Mn--Ni, Ni--Si and Mn--Ni--Si systems, the mean \(S_{\rm B2}\) of these random states was within \num{0.012} of zero, with a standard deviation of \num{0.05}--\num{0.06}.

At 1~at.\% Ni, the values were first averaged over the 20 sampled configurations from each simulation. Across the ten simulation averages, \(\rho_{\rm 1NN}\) is \(-0.18\pm0.01\) for Mn--Ni, \(-0.39\pm0.01\) for Ni--Si and \(-0.49\pm0.02\) for Mn--Ni--Si. The corresponding \(\rho_{\rm 2NN}\) values are \(0.08\pm0.03\), \(0.18\pm0.04\) and \(0.53\pm0.03\). These values give \(S_{\rm B2}\) of about \num{0.13}, \num{0.29} and \num{0.51}, respectively. The \(\pm\) values are standard deviations across the ten simulation averages. For the Mn--Ni--Si interface, the Ni--Ni bond ratio increases from below \num{1} at 1NN to about \num{1.4} at 2NN. The same reversal occurs at all four nonzero Ni concentrations.

We use this local measure instead of a bulk long-range order parameter because the ordered atoms form a finite group attached to the CRP rather than a large ordered phase.

The additional energy reduction \(\Delta E_{\rm add}^{X}\) is defined in Section~2.5 of the main text, where \(X\) denotes Mn, Si or Mn--Si. The four endpoint energies were obtained from independent simulations using the same cells with \num{27000} sites. A negative value means that \(X\) gives an additional energy reduction beyond the paired change for the Ni-only system at the same total Ni concentration. Because \(\Delta E_{\rm add}^{X}\) uses formation energies for the complete simulation cell, it includes energy changes throughout the cell.

\section{Effect of structural vacancies on the evolution from shell to appendage}
\label{si:morphvacancy}

The shell-to-appendage evolution is compared with and without structural vacancies. The change from a shell to an appendage in Fig.~2 of the main text was obtained from the \(60\times60\times60\) canonical trajectory for Fe--0.2Cu--1.5Ni--1.5Mn--0.5Si--0.1P at.\% without vacancies \cite{SI:Ke2025Vacancies}. A separate calculation examined vacancies within the population rich in Mn, Ni and Si. Our previous work using the same cluster expansion showed how these vacancies stabilize the clusters and change their ordering pathway \cite{SI:Ke2025Vacancies}. The calculation used the Ni\(_{16}\)Mn\(_6\)Si\(_7\)Va\(_3\) site ratio from the bcc G-phase mapping \cite{SI:whiting2020gamma2}. The site fractions were \(x_{\rm Ni}=0.500\), \(x_{\rm Mn}=0.1875\), \(x_{\rm Si}=0.21875\) and \(x_{\rm Va}=0.09375\). Compared with the trajectory without vacancies, this calculation shows a larger displacement of the region rich in Cu and a more distinct appendage. No closed core--shell stage appears among the sampled configurations.

\begin{figure}[H]
\centering
\includegraphics[width=0.85\textwidth]{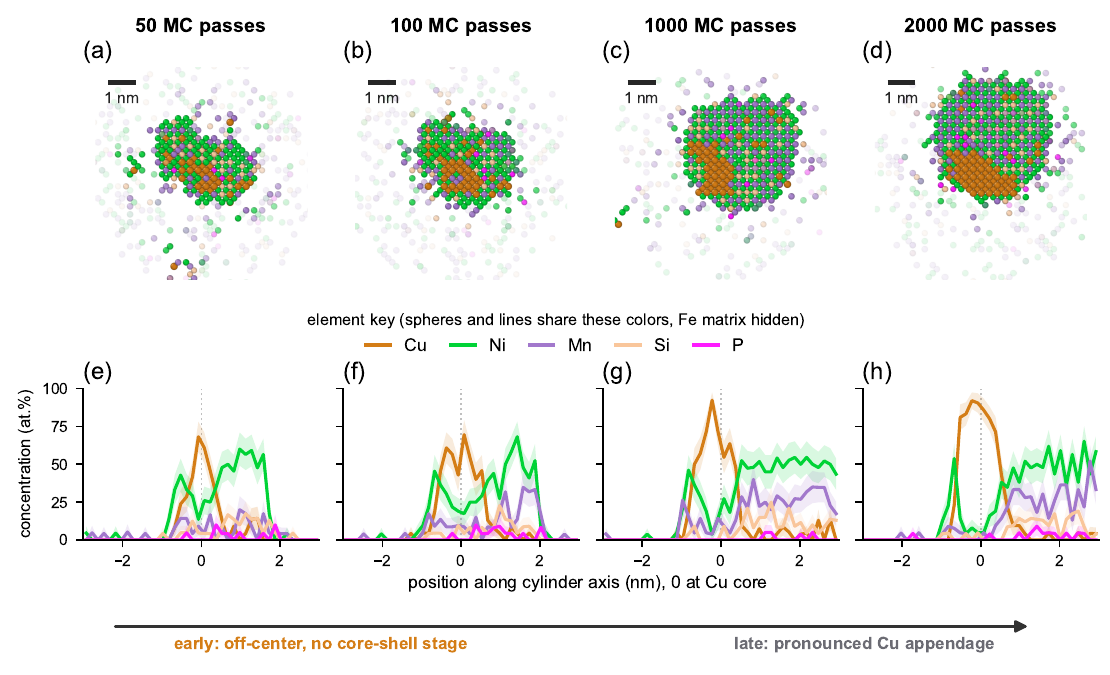}
\caption{CRP--MNSP evolution with structural vacancies at \SI{573}{\kelvin}. The sequence without vacancies is shown in Fig.~2 of the main text. The population of Mn, Ni, Si and vacancies follows the Ni\(_{16}\)Mn\(_6\)Si\(_7\)Va\(_3\) ratio of the bcc G-phase mapping. Vacancies are hidden and the configurations use the same MC pass indices as Fig.~2.}
\label{fig:morphologyvacancy}
\end{figure}

\section{Dependence of CRP--MNSP morphology on the amount of Mn, Ni and Si relative to Cu}
\label{si:seeded}
 We performed seeded \(30\times30\times30\) simulations to see how the CRP--MNSP morphology depends on the amount of Mn, Ni and Si relative to Cu. The large-cell trajectory in Fig.~2 of the main text suggests that the morphology changes as the MNSP grows relative to the Cu-rich region. Independent calculations were performed in \(30\times30\times30\) cells at \SI{573}{\kelvin}. The Cu concentration was fixed at \num{0.5}~at.\%. The Mn, Ni and Si concentrations were varied together at fixed atomic proportions of 3 Ni to 3 Mn to 1 Si. Their total amount relative to Cu was described by
\begin{equation}
R_{\rm MNS/Cu}=\frac{x_{\rm Mn}+x_{\rm Ni}+x_{\rm Si}}{x_{\rm Cu}}.
\end{equation}
The tested ratios correspond to Ni concentrations from \num{0.43} to \num{3.0}~at.\%. At each ratio, three independent calculations started from a random solid solution, a Cu core enclosed by an Mn--Ni--Si shell or a Cu precipitate with the Mn--Ni--Si atoms placed on one side as an attached Mn--Ni--Si region. The discrete ratios are shown in \cref{fig:seededconvergence}.

The morphology was measured using the normalized Cu offset \(d_{\rm Cu}/r_{\rm MNSP}\), where \(d_{\rm Cu}\) is the Cu-to-MNSP center distance and \(r_{\rm MNSP}\) is the MNSP equivalent radius (Fig.~3a of the main text). For each \num{12000}-pass trajectory, the offset was averaged over the second half of the calculation. The error bars in \cref{fig:seededconvergence} give one standard deviation over the same interval. We used an offset of \num{0.45} to separate the compact configurations from the appendages near \num{0.71}. This value only guides the comparison among the simulated configurations and does not define a thermodynamic transition or an equilibrium contact angle. For all three initial states, the final offsets remain below this value through \(R_{\rm MNS/Cu}=7\), which corresponds to \num{1.5}~at.\% Ni. At \(R_{\rm MNS/Cu}=14\), which corresponds to \num{3.0}~at.\% Ni, all three offsets are near \num{0.71}. The Cu precipitate is therefore close to the edge of a compact Mn--Ni--Si domain. The morphology changes within the tested interval \(7<R_{\rm MNS/Cu}<14\).

The three initial states reach the same morphology class at each composition ratio. For \(R_{\rm MNS/Cu}\leq7\), the initial appendage contracts toward the Cu core. At \(R_{\rm MNS/Cu}=14\), the initial shell moves outward and forms an appendage. The agreement among the three initial states shows that the final morphology depends on the amount of Mn, Ni and Si relative to Cu and is not set by the initial geometry. This comparison applies to the simulated finite particles and does not give an equilibrium contact angle or a physical transformation rate. The initial and final configurations are compared in \cref{fig:seededmontage}.

\begin{figure}[H]
\centering
\includegraphics[width=0.55\textwidth]{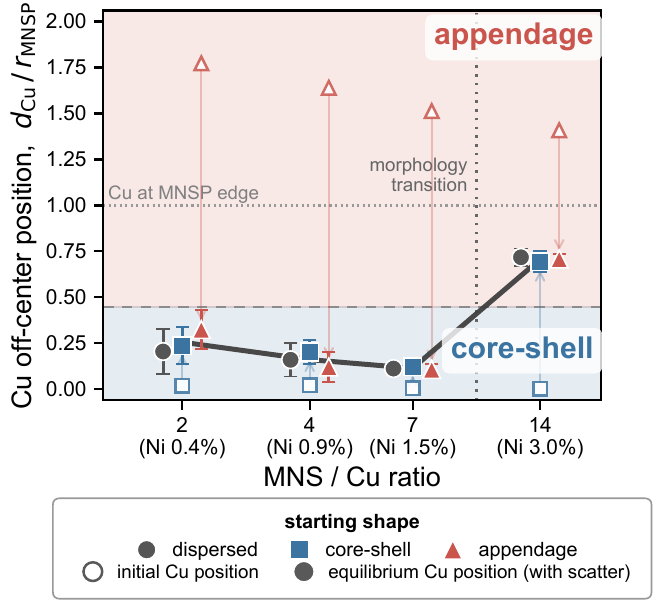}
\caption{CRP--MNSP morphology as a function of \(R_{\rm MNS/Cu}\) at \SI{573}{\kelvin} measured by \(d_{\rm Cu}/r_{\rm MNSP}\). Cu is fixed at \num{0.5}~at.\% and the fixed atomic proportions are 3 Ni to 3 Mn to 1 Si. The three initial configurations give core--shell particles for \(R_{\rm MNS/Cu}\leq7\) and appendages at \(R_{\rm MNS/Cu}=14\).}
\label{fig:seededconvergence}
\end{figure}

\begin{figure}[H]
\centering
\includegraphics[width=0.85\textwidth]{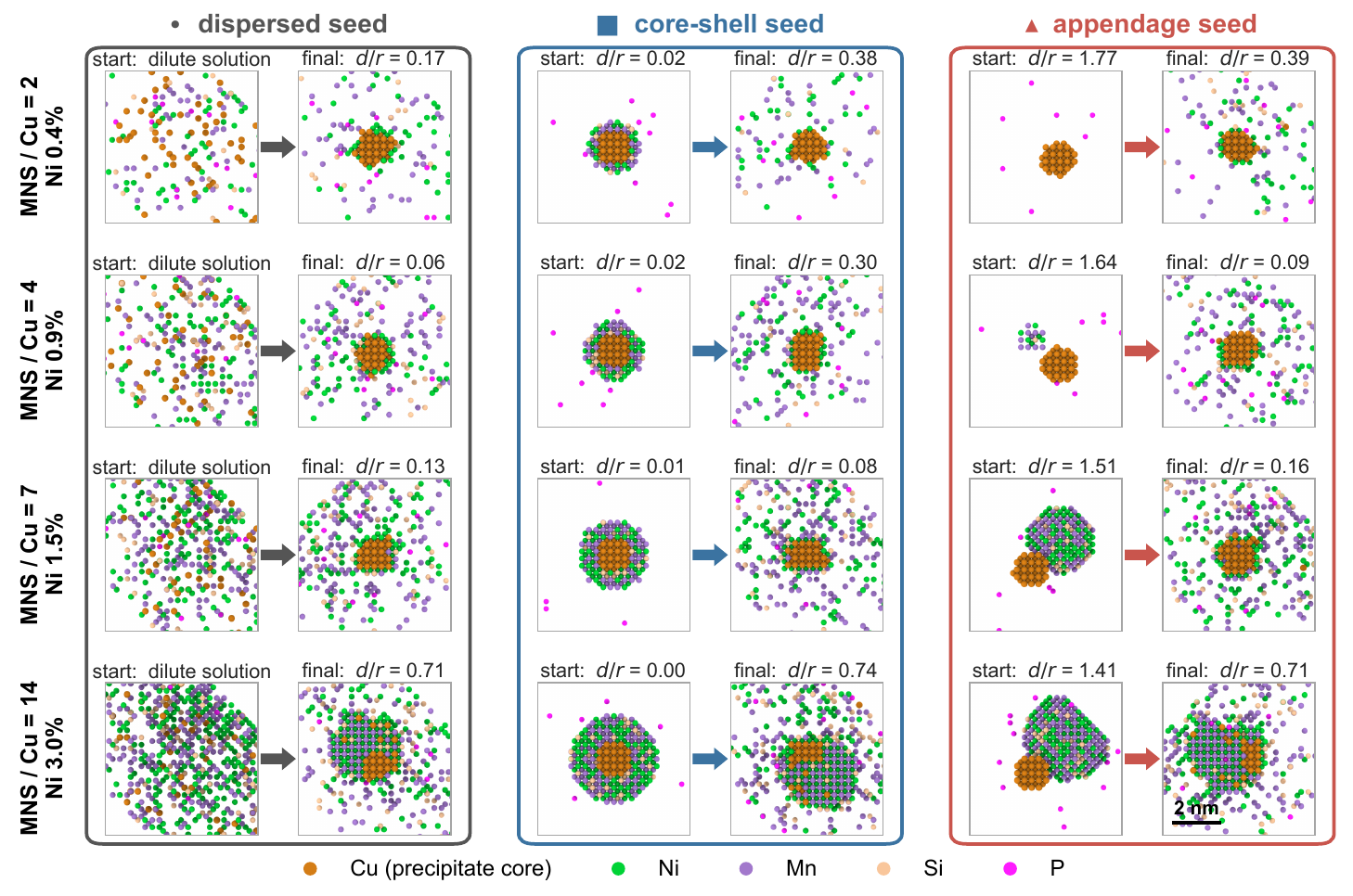}
\caption{Initial and final configurations for the composition sweep in \cref{fig:seededconvergence}. The rows correspond to increasing \(R_{\rm MNS/Cu}\). The columns give the three initial configurations and their final states, labeled by \(d/r=d_{\rm Cu}/r_{\rm MNSP}\). The Fe matrix is hidden.}
\label{fig:seededmontage}
\end{figure}

\section{McLean linearization of the isotherms for individual solutes}
\label{si:mclean}

To test whether the isotherms for individual solutes follow the ideal Langmuir--McLean form, the fits used in the main text were linearized at \(r_{\rm eq}=\SI{1.0}{\nano\metre}\) \cite{SI:langmuir1918adsorption,SI:mclean1957grain,SI:hondros1977theory}. The axes in \cref{fig:mclean} are \(\ln[\theta/(1-\theta)]\) and \(\ln[x/(1-x)]\), where \(\theta=\Gamma_i/\Gamma_{\rm sat}\). For the equivalent surface area of the \SI{1.0}{\nano\metre} CRP, the common value \(\Gamma_{\rm sat}=\SI{13.3}{atoms.nm^{-2}}\) corresponds to approximately 167 sites. We used this value to normalize the coverage and was not fitted separately for each solute. For the ideal form with one type of site, the slope is one and the intercept is \(\ln K\), with \(\Delta G_{\rm seg}=-k_{\rm B}T\ln K\).

These linearized fits use the same fixed-radius composition sweep as Fig.~9a of the main text. With the slope fixed to one, the Ni fit gives \(\Delta G_{\rm seg}=\SI{-297}{\milli\electronvolt}\), the same value reported there. Differences from the fitted line appear mainly at the highest coverages, where \(\theta=\Gamma_{\rm Ni}/\Gamma_{\rm sat}\) approaches one. The Mn isotherm again does not give a single composition-independent \(\Delta G_{\rm seg}\). The \SI{-74}{\milli\electronvolt} value is reported only as an effective estimate over the range shown. Si is depleted in the single-solute calculations and is not fitted with the McLean form.

\begin{figure}[t]
\centering
\includegraphics[width=0.45\textwidth]{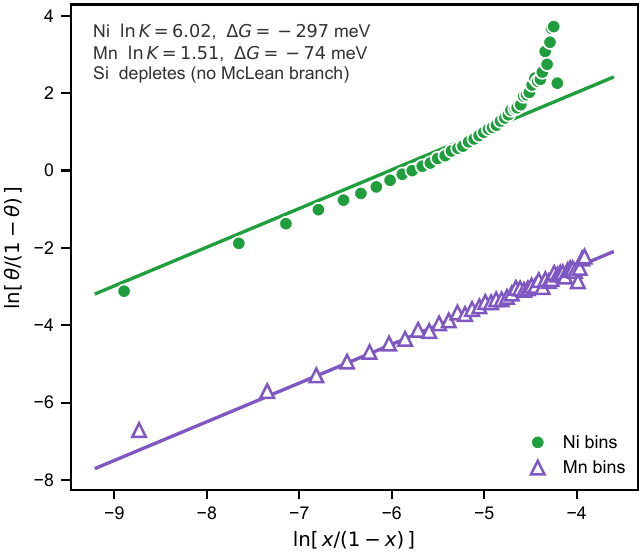}
\caption{McLean linearization of the isotherms for individual solutes at \SI{573}{\kelvin} and \(r_{\rm eq}=\SI{1.0}{\nano\metre}\), using the fixed-radius composition sweep and the fixed-slope fit used for Fig.~9a of the main text. The Ni fit with the slope fixed to one gives \(\Delta G_{\rm seg}=\SI{-297}{\milli\electronvolt}\), the same fixed-slope fit reported there. Ni deviates from the fitted line at the highest coverages shown here (\(\theta\to1\)). Mn deviates from a slope of one. The \SI{-74}{\milli\electronvolt} value is reported only as an effective estimate. Si is depleted and is not fitted.}
\label{fig:mclean}
\end{figure}

\section{Semigrand-canonical calculation of the Fe/CRP free-energy changes caused by Ni, Mn and Si}
\label{si:gibbssgc}

We combined interfacial excesses from a fixed-radius canonical alloying scan with the relation between matrix composition and diffusion potential from semigrand-canonical (SGC) calculations \cite{SI:gibbs1878equilibrium,SI:cahn1979thermodynamics,SI:kaplan2013review} to obtain the relative changes in Fe/CRP interfacial free energy reported in the main text. These thermodynamic paths differ from the canonical growth calculation, which gives a change in internal energy at fixed composition. The SGC calculations used the same multicomponent CE and \(20\times20\times20\) cell described in Section~\ref{si:methods}, with four forward and four reverse scans for each species over diffusion-potential windows that contain the target dilute compositions. Ni, Mn and Si were evaluated along three separate single-solute paths. The interfacial excess \(\Gamma_i\) for each species came from a canonical-ensemble alloying scan on a single Cu-rich precipitate with \(r_{\rm eq}\approx\SI{1.0}{\nano\metre}\), embedded in the Fe matrix, in which the matrix content of one species at a time was raised from zero in several independent simulations.

\Cref{fig:gibbssgc}a shows the matrix content of each species as a function of its own diffusion potential, relative to a dilute reference state. \Cref{fig:gibbssgc}b shows the corresponding interfacial excess \(\Gamma_i\) along the same paths. Integrating the Gibbs adsorption relation along each curve gives the change in Fe/CRP free energy in \cref{fig:gibbssgc}c. Starting from a near-zero matrix composition and integrating up to \num{1}~at.\% in the matrix gives \(\Delta\gamma_{\rm Fe/CRP}\)=\SI{-0.150}{\joule\per\square\metre} for Ni (\(\Gamma_{\rm Ni}\)=\SI{10.72}{atoms.nm^{-2}}), \SI{-0.004}{\joule\per\square\metre} for Mn (\(\Gamma_{\rm Mn}\)=\SI{0.58}{atoms.nm^{-2}}) and \SI{+0.002}{\joule\per\square\metre} for Si (\(\Gamma_{\rm Si}\)=\SI{-0.21}{atoms.nm^{-2}}). These values represent the single-radius Gibbs result used in the main text (Section~3.3).

\begin{figure}[t]
\centering
\includegraphics[width=0.85\textwidth]{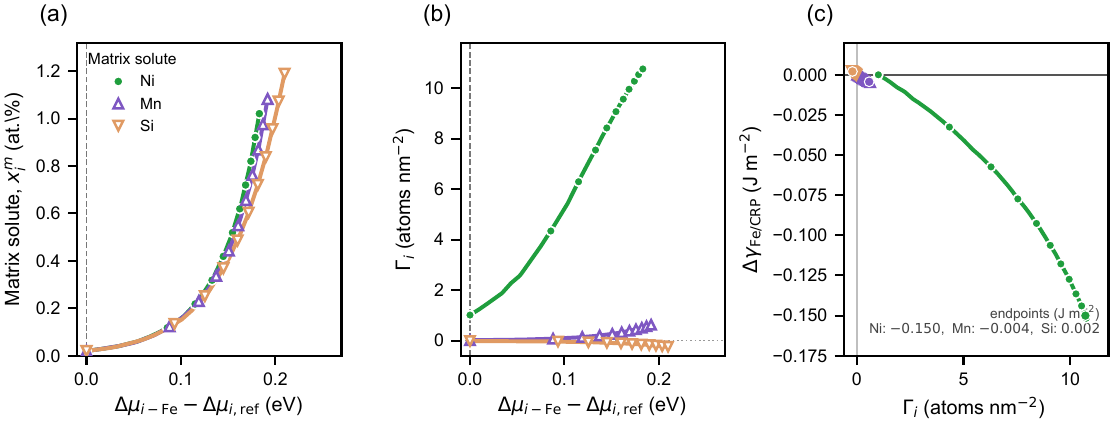}
\caption{Semigrand-canonical calculation of the Fe/CRP free-energy changes along separate Ni, Mn and Si paths at \SI{573}{\kelvin}. (a)~Matrix content of each species as a function of its own diffusion potential, relative to a dilute reference state. (b)~Interfacial excess \(\Gamma_i\) for each species over the same interval. (c)~Change in Fe/CRP free energy relative to the near-zero-composition state, as a function of \(\Gamma_i\), with each species' endpoint value at \num{1}~at.\% matrix content labeled.}
\label{fig:gibbssgc}
\end{figure}

\section{Interaction analysis of binary co-segregation}
\label{si:binary}

To describe the binary co-segregation discussed in the main text, we related the change in the segregation free energy of solute \(i\) to the interfacial coverage of solute \(j\) \cite{SI:guttmann1975equilibrium,SI:guttmann1982thermodynamics}.
\begin{equation}
\Delta G_{\rm seg,i}-\Delta G_{\rm seg,i}^{0}
=\alpha_{ij}^{\rm eff}Y_j,
\qquad
Y_j=\Gamma_j/\Gamma_{\rm sat}.
\label{eq:si_coseg}
\end{equation}

Here, \(\Delta G_{\rm seg,i}^{0}\) is the reference with one solute and \(\alpha_{ij}^{\rm eff}\) is the effective interaction coefficient at the Fe/CRP interface. A negative coefficient denotes attractive co-segregation. The coefficients are \SI{-7.1}{\kilo\joule\per\mole} for Mn segregation with Ni and \SI{-7.8}{\kilo\joule\per\mole} for Ni segregation with Mn. Ni segregation with Si gives a more negative coefficient of \SI{-30.8}{\kilo\joule\per\mole} (\cref{fig:coseg}a).

The two Mn data points at the lowest Ni coverage are slightly positive because the Mn reference with no other solute changes with matrix composition. When each simulation is instead referenced to the state without Ni and with the same matrix composition, resampling of the ten states gives a Mn--Ni slope between about \SI{-8.2}{} and \SI{-5.8}{\kilo\joule\per\mole}. The Mn--Ni interaction therefore remains weakly attractive after this correction.

The excess comparison in \cref{fig:coseg}b uses the corresponding state without Ni as the reference for each chemistry. The Mn excess increases from a positive baseline with a slope of \num{+0.061}. The Si excess increases more strongly with a slope of \num{+0.128} and changes from depletion to enrichment near \(\Gamma_{\rm Ni}=\SI{0.9}{atoms.nm^{-2}}\). The P excess changes little and has a slope of \num{+0.011}. Ni therefore increases the Mn and Si excesses but has little effect on P.

We also compared the co-segregation over the common range of particle sizes using references with one solute adjusted to the measured matrix composition and the CRP radius of each binary state. Ni--Si gives the strongest co-segregation, while Mn--Ni gives a smaller effect and P remains weak. The same relative strengths are observed throughout the common size interval and are not limited to the \(r_{\rm eq}=\SI{1.0}{\nano\metre}\) comparison in the main text.

\begin{figure}[t]
\centering
\includegraphics[width=0.75\textwidth]{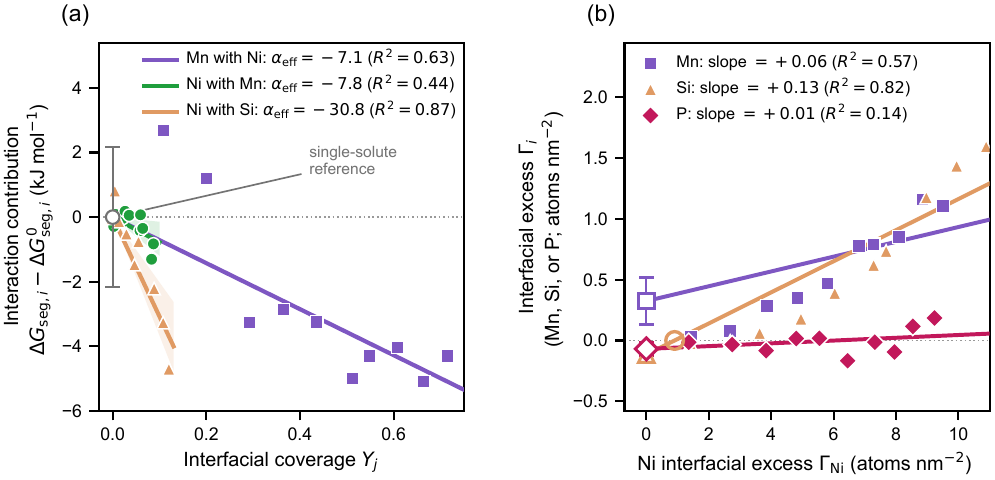}
\caption{Binary co-segregation around Ni at \SI{573}{\kelvin}. (a)~Interaction contribution as a function of the coverage by the second solute. (b)~Mn, Si and P interfacial excesses as a function of Ni excess.}
\label{fig:coseg}
\end{figure}

\section{Wetting sensitivity to the B2 reference}
\label{si:b2wetting}

The main text uses the bcc mapping of Ni\(_{16}\)Mn\(_6\)Si\(_7\)Va\(_3\) as the MNSP reference because it preserves the G-phase composition and structural vacancies \cite{SI:whiting2020gamma2}. As a second reference, we analyzed stoichiometric NiMn B2 with Ni and Mn on opposite bcc sublattices, using the same cluster-expansion energy. This B2 state is relevant because order similar to B2 appears in the large-cell trajectory without vacancies.

At \SI{0}{\kelvin}, the Fe/B2 coefficient ranges from \SI{0.329}{} to \SI{0.332}{\joule\per\square\metre} over \(r_{\rm eq}=\SIrange{0.656}{1.780}{\nano\metre}\), and the Cu/B2 coefficient ranges from \SI{0.237}{} to \SI{0.242}{\joule\per\square\metre} (\cref{fig:b2wetting}a). Combining these values with the \SIrange{0.50}{0.52}{\joule\per\square\metre} Fe/CRP coefficient at \SI{573}{\kelvin} gives \(S_E\) between \SI{-0.074}{} and \SI{-0.045}{\joule\per\square\metre}. The corresponding \(W_{\mathrm{adh},E}\) is between \SI{0.586}{} and \SI{0.615}{\joule\per\square\metre}.

With the MNSP(G-phase) reference, \(S_E\) ranges from  \SI{-0.48}{} to \SI{-0.30}{\joule\per\square\metre} and \(W_{\mathrm{adh},E}\) ranges from  \SI{0.66}{} to \SI{0.84}{\joule\per\square\metre}. At the common radius used in the main text, \(S_E\approx\SI{-0.40}{\joule\per\square\metre}\) and \(W_{\mathrm{adh},E}=\SIrange{0.72}{0.79}{\joule\per\square\metre}\) (\cref{fig:b2wetting}b). Both references combine the ordered phase coefficients at \SI{0}{\kelvin} with the Fe/CRP coefficient at \SI{573}{\kelvin}. Note that the comparison describes the balance based on internal energy and not equilibrium wetting.

Within this mixed-temperature internal-energy comparison, both references give a positive work of adhesion and a negative spreading parameter, consistent with attachment and partial coverage. The NiMn B2 spreading parameter is much closer to \(S_E=0\) than the MNSP(G-phase) value. This difference is consistent with the nearly closed shell in the trajectory without vacancies and the larger displacement from the center in \cref{fig:morphologyvacancy}. Because the two phases also differ in composition and sublattice chemistry, this comparison does not isolate the vacancy contribution. Section~\ref{si:mnspcrosscheck} gives a corresponding independent comparison for the MNSP(G-phase) coefficients themselves, using flat-slab values instead of an alternative ordered phase.

\begin{figure}[htbp]
\centering
\includegraphics[width=0.95\textwidth]{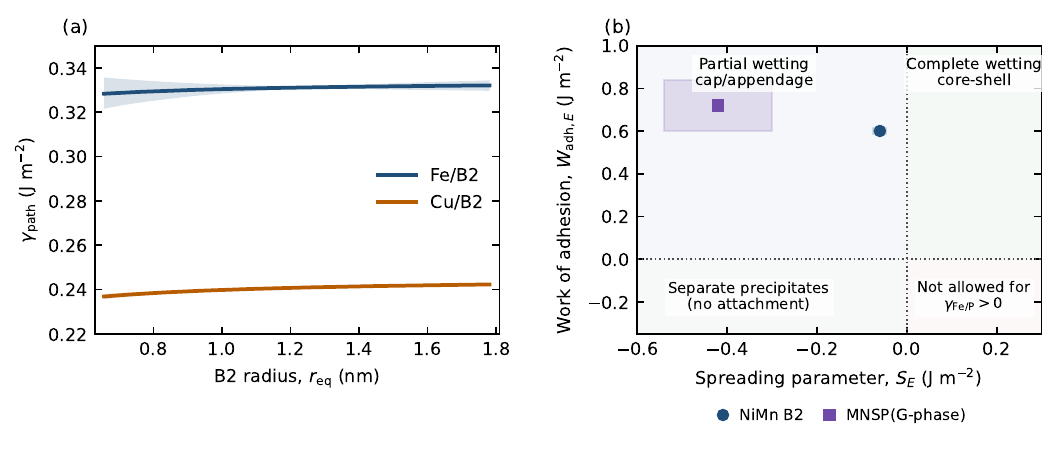}
\caption{Effect of the ordered phase reference on the interface balance. (a)~Fe/B2 and Cu/B2 coefficients \(\gamma_{\rm path}\) at \SI{0}{\kelvin}. (b)~Balances from the ordered phase coefficients at \SI{0}{\kelvin} and the Fe/CRP coefficient at \SI{573}{\kelvin}. Boxes give the bounds from the complete coefficient ranges.}
\label{fig:b2wetting}
\end{figure}

\section*{References for the Supplementary Information}

\endgroup
\end{document}